\documentclass[a4paper,11pt]{article}
\pdfoutput=1 

\usepackage{jheppub} 
\usepackage[utf8]{inputenc}

\usepackage{setspace}

\definecolor{refkey}{gray}{0.45}
\definecolor{labelkey}{RGB}{155,48,48}

\usepackage{physics}
\usepackage[obeyFinal]{todonotes}

\def\beq{\begin{eqnarray}}\def\eeq{\end{eqnarray}}
\def\be{\begin{equation}}\def\ee{\end{equation}}

\def\mes[#1]{d^{3}{#1}}

\def\del{\partial}

\newcommand{\half}{\frac{1}{2}}
\newcommand{\quarter}{\frac{1}{4}}

\def\si{\sigma}

\def\del{\partial}
\newcommand{\ads}[1]{AdS$_{#1}$}

\reversemarginpar

\title{On the Dynamics of Near-Extremal Black Holes}

 \author[a,b]{Pranjal Nayak,}
 \author[a,c]{Ashish Shukla,}
 \author[a,d]{Ronak M Soni,}
 \author[a]{Sandip P. Trivedi,}
 \author[a]{V. Vishal}
 \affiliation[a]{\it Department of Theoretical Physics,
 Tata Institute of Fundamental Research,\\  Colaba, Mumbai, India, 400005\\}
 \affiliation[b]{\it Department of Physics \& Astronomy,
 University of Kentucky,\\  Lexington, Kentucky, USA, 40506-0055\\}
 \affiliation[c]{\it Department of Physics and Astronomy,
 University of Victoria,\\  Victoria, British Columbia, Canada, V8P 5C2\\}
 \affiliation[d]{\it Perimeter Institute for Theoretical Physics,\\
 Waterloo, Ontario, Canada, N2L 2Y5\\}

\emailAdd{nayak.pranjal@gmail.com}
\emailAdd{ashish@uvic.ca}
\emailAdd{ronak@theory.tifr.res.in}
\emailAdd{sandip@theory.tifr.res.in}
\emailAdd{vishal@theory.tifr.res.in}

\vspace{1cm}

\abstract{We analyse the dynamics of near-extremal Reissner-Nordstr{\"o}m black holes in asymptotically four-dimensional Anti de Sitter space (AdS$_4$). We work in the spherically symmetric approximation and study the thermodynamics and the response to a probe scalar field. We find that the behaviour of the system, at low  energies and to leading order in our approximations,  is well described by the Jackiw-Teitelboim (JT) model of gravity.  In fact, this behaviour can be understood from symmetry considerations and arises due to the breaking of time reparametrisation invariance. The JT model has been analysed in considerable detail recently and related to the behaviour of the SYK model. Our results indicate that  features in  these models which arise  from symmetry considerations alone are   more general and present quite universally in near-extremal black holes.}

\preprint{\parbox{3cm}{TIFR/TH/17-35}}

\begin{document}
\maketitle
\flushbottom

\section{Introduction }

Recently, a new class of solvable  quantum mechanical models have gained considerable attention. 
The study of these models was initiated by Sachdev and Ye, \cite{Sachdev:1992fk}, and Kitaev, \cite{Kitaev-talks:2015}. Subsequently these and similar models have been discussed extensively,  a partial set of references is \cite{Jevicki:2016bwu,Maldacena:2016hyu,Jevicki:2016ito,Gu:2016oyy,Gross:2016kjj,Berkooz:2016cvq,Garcia-Garcia:2016mno,Fu:2016vas,Witten:2016iux,Gurau:2016lzk,Klebanov:2016xxf,Davison:2016ngz,Krishnan:2016bvg,Li:2017hdt,Murugan:2017eto,Gross:2017vhb,Bulycheva:2017uqj,Gaikwad:2017odv,Eberlein:2017wah,Yoon:2017nig,Sonner:2017hxc,Choudhury:2017tax,Krishnan:2017lra,Bhattacharya:2017vaz,Forste:2017apw}. We will refer to this class of models as the SYK models below. 
These models  have the virtue of being simple enough to be exactly solvable in the large $N$ limit, yet being rich enough to possess some highly non-trivial features of interacting systems. Most notably, real time correlations in these models are found to thermalise, and out of time order correlations were found to saturate a bound that governs the onset of chaotic behaviour.

These developments are of considerable interest from the point of view of studying black hole dynamics, \cite{Hayden:2007cs,Sekino:2008he}. It has been shown that quite generally the chaos bound is satisfied by black holes \cite{Shenker:2014cwa}. 
A long standing puzzle in the study of black holes pertains to the  information loss issue.
The SYK models share some highly non-trivial features in common with black holes. In particular, as mentioned above, they  saturate the chaos bound; yet these models are consistent quantum mechanical theories which should not lead to information loss.\footnote{The original SYK model involves random couplings which must be averaged over. However,  similar models can be formulated which are manifestly unitary without such an averaging over couplings, see \cite{Witten:2016iux,Klebanov:2016xxf}.}  One would hope that a more detailed analysis of these models could therefore help shed light on how the information loss puzzle is resolved for black holes. 

In particular, one might hope that the SYK models are  related to the study of near-extremal black holes, and quite a bit of recent investigation in \ads{2}/CFT$_1$ correspondence has been directed to improve our understanding of the same, \cite{Castro:2008ms,Castro:2014ima,Jensen:2016pah,Maldacena:2016upp,Engelsoy:2016xyb,Almheiri:2016fws,Cvetic:2016eiv,Mandal:2017thl,Gross:2017hcz,Forste:2017kwy,Das:2017pif,Kyono:2017pxs,Eling:2017txo,Dartois:2017xoe,Krishnan:2017txw,Dubovsky:2017cnj,Caputa:2017yrh,Taylor:2017dly,Grumiller:2017qao,deBoer:2017xdk,Jian:2017tzg,Cai:2017nwk,Halmagyi:2017leq,Kitaev:2017awl,Das:2017hrt,Das:2017wae,Haehl:2017pak,Mertens:2018fds}. It was found that the low-energy behaviour of the SYK model is characterised by the emergence of a local conformal symmetry consisting of time-reparametrisations. This symmetry is explicitly broken since the UV degrees of freedom do not fully decouple, and is also spontaneously broken by the ground state.  
The resulting low-energy modes, which are the analogues of Goldstone modes, are governed by an action determined by symmetry considerations alone and their behaviour  governs the low-energy dynamics of the system. In particular, it gives rise to a linear specific heat and also to out of time order correlation functions which saturate the chaos bound. This action, which involves a Schwarzian derivative of the time-reparametrisations, will be referred to as the Schwarzian action below. 

It was found by \cite{Achucarro:1993fd,Jensen:2016pah,Maldacena:2016upp,Engelsoy:2016xyb} that many of these properties are  in fact true in a two-dimensional theory of gravity first studied by Jackiw and Teitelboim \cite{Teitelboim:1983ux,JACKIW1985343}. The  time-reparametrisation modes do arise in this model and are governed  by the Schwarzian action which  gives rise to a linear specific heat and the saturation of the chaos bound. We will refer to this system as the JT model below. 

Near-extremal black holes are known to share some properties with the JT model.  For example, their near-horizon geometry is well known to involve two-dimensional Anti-de Sitter space, AdS$_2$, whose asymptotic symmetries are  the time-reparametrisations referred to above. It is also well-known that analysing the excitations above extremality for these black holes requires one to retain more than the near-horizon AdS$_2$ region, leading to the explicit breaking of this symmetry.  
These similarities suggest that the lessons learnt from the study of SYK models could apply more generally for the study of these black holes.

This paper is devoted to studying this issue in more detail. 
 In particular, we  study  spherically symmetric   near-extremal black holes  in higher dimensions  and analyse their dynamics in the spherically symmetric ($S$-wave)  sector. 
For concreteness  we  restrict ourselves to  near-extremal Reissner-Nordstr{\"o}m black holes in asymptotically AdS$_4$ spacetimes, and consider the simplest system, gravity coupled to a Maxwell field, where they arise. 

Our main results are  that the thermodynamics and the low-energy behaviour of this system   is in fact well approximated by the JT model, and exhibits the breaking of time reparametrisation symmetry with a Schwarzian  action. We analyse the low-energy behaviour of the system by coupling it to a probe scalar field and calculate the four-point function at zero temperature.
We find that this is well approximated, at leading order, by  coupling the time reparametrisation modes to the scalar field in a manner determined by symmetry considerations, as happens in the JT model.

The paper is organised as follows. In section 2, we introduce the system of interest,  consisting of Einstein gravity and the Maxwell field and analyse some of its properties, including  near-extremal thermodynamics.
 In section 3, we introduce additional matter in the form of a scalar field, analyse the resulting $4$-point scalar correlator in the four-dimensional asymptotically AdS$_4$ system, and take its low energy limit. In section 4, we consider the  JT model and show that,  for a suitable choice of parameters, it reproduces the near-extremal thermodynamics, as well as the four-point function at low energies and zero temperature.
 In section 5, we carry out an S-wave reduction of the Einstein-Maxwell system and show why it agrees with the JT model to leading order at low-energies. We end in section 6 with conclusions and some future directions. 
 Appendices \ref{fullthermo}, \ref{genonact}, \ref{4dto2d}, \ref{thermoap}, \ref{secoup}, \ref{dimredsource} contain important supplementary material. 
 
 Before proceeding let us mention some other important references. A thorough study of the JT model, including the first computation of the four-point function, was done by \cite{Almheiri:2014cka}. Also, the fact that the  JT model correctly captures the near-extremal thermodynamics was noted by \cite{Maldacena:2016upp,Almheiri:2016fws}.
 An S-wave dimensional reduction from Einstein-Chern-Simons theory in AdS$_3$ was used to get the JT model with a gauge field in \cite{Gaikwad:2018}. 
 Also, \cite{Cvetic:2016eiv}  has elucidated various possible solutions in a 2D gravity model with a dilaton and a gauge field. Relevant works examining dimensional reduction and near-horizon physics include \cite{Boonstra:1998mp,Skenderis:2006uy,Gouteraux:2011qh}.

\section{Spherically Symmetric Reissner-Nordstr{\"o}m Black Holes}
\label{RNBH}
In this section we analyse spherical charged black holes that arise in a theory of gravity coupled to a Maxwell field in the presence of a negative cosmological constant.
The action is 
\begin{equation}
\label{act1}
S={1\over 16 \pi G}\int d^4 x\sqrt{-g}\,\big(R-2\Lambda\big)\,-\,\frac{1}{4G}\int d^4 x \sqrt{-g}\,F_{\mu\nu}F^{\mu\nu}.
\end{equation}
This system  is well known to have the Reissner-Nordstr{\"o}m black hole solution given by
\begin{align}
\label{RNAdS4II}
ds^2&=-a(r)^2\,dt^2\,+\,a(r)^{-2}\,dr^2\,+\,b(r)^2\,(d\theta^2\,+\,\text{sin}^2\theta \,d\varphi^2),\nonumber\\
a(r)^2&=1-\frac{2GM}{r}+\frac{4\pi(Q_m^2+Q_e^2)}{r^2}+\frac{r^2}{L^2},\\
b(r)^2&=r^2,\nonumber\\[10pt]
F_{rt}&=\frac{Q_e}{r^2},\label{Qe}\\
F_{\theta\varphi}&=Q_m\, \text{sin} \, \theta\label{Qm}.
\end{align}
Here $M, Q_e, Q_m$ are the mass, the electric and the magnetic charges of the black hole, and  $L$ is the AdS$_4$ radius,
\begin{equation}
\label{valL}
L=\sqrt{\frac{3}{|\Lambda|}}.
\end{equation}
This solution  is manifestly spherically symmetric and preserves the $SO(3)$ rotational symmetry. 

An electromagnetic duality transformation under which
\begin{equation}
\label{emdulity}
\left ( \begin{array}{c}
Q_m \\
Q_e \end{array} \right) \rightarrow \left(\begin{array}{cc} \cos\chi & \sin\chi \\ -\sin\chi & \cos\chi \end{array}\right) \left (\begin{array}{c} Q_m \\ Q_e \end{array}\right),
\end{equation}
allows one to map the general solution with both electric and magnetic charges to the purely magnetic case, where $Q_e=0$, and keeps the metric, eq.(\ref{RNAdS4II}), invariant. 
We will use this  duality transformation to work with the purely magnetic case below. 

It is easy to see that in the asymptotic region,  $r\rightarrow \infty$, the metric, eq.(\ref{RNAdS4II}),  becomes  AdS$_4$,
\begin{equation}
\label{ads4}
ds^2=-\pqty{1+\frac{r^2}{L^2}}\,dt^2\,+\,\pqty{1+\frac{r^2}{L^2}}^{-1}dr^2\,+\,r^2\,(d\theta^2\,+\,\text{sin}^2\theta \,d\varphi^2).
\end{equation}

At extremality,
\begin{align}
Q_m^2 &= \frac1{4\pi} \pqty{r_h^2+\frac{3r_h^4}{L^2}},\label{Qext}\\
M_{ext}&= \frac{r_h}{G} \pqty{1+\frac{2r_h^2}{L^2}}\label{Mext},\\[10pt]
a^2(r) &= \frac{ (r-r_h)^2 }{ r^2 L^2 } \, \pqty{ L^2 + 3 r_h^2 + 2 r r_h + r^2 },\label{axt}
\end{align}
and the temperature of the black hole vanishes. 
To simplify the discussion we focus henceforth on big black holes where the horizon size is much bigger than the AdS radius, 
\begin{equation}
\label{bigb}
r_h \gg L.
\end{equation}
The charge and mass are then given in terms of the horizon radius $r_h$ by  
\begin{align}
Q_m^2 & \simeq \frac1{4\pi} \frac{3r_h^4}{L^2},\label{Qext2}\\
M_{ext}&\simeq\frac{2r_h^3}{G L^2}\label{Mext2}.
\end{align}
The near-horizon region of the extremal geometry,  $r-r_h\ll r_h$,  has the metric, upto $O({r-r_h\over r_h})$ corrections,
\begin{equation}
\label{nearhor}
ds^2= \bqty{ -\frac{(r-r_h)^2}{L_2^2}dt^2\,+\,\frac{L_2^2}{(r-r_h)^2}dr^2\,+\,r_h^2\,(d\theta^2\,+\,\text{sin}^2\theta \,d\varphi^2) }.
\end{equation}
This is of the form AdS$_2\times$S$^2$, where the radius of the AdS$_2$ and the $S^2$ are given by
\begin{align}
R_{AdS_2}&=L_2\simeq\frac{L }{\sqrt{6}},\label{ads2rad}\\
R_{S^2}&=r_h.\label{S2rad}
\end{align}
The asymptotic region of the AdS$_2$ geometry is given by 
\be
\label{csta}
{r-r_h\over L_2}\gg 1.
\ee

Eq.(\ref{nearhor}) is a good approximation when 
\begin{equation}
\label{condaar}
{r-r_h\over r_h}\ll 1.
\end{equation}
This is consistent with eq.(\ref{csta}) for a big black hole meeting eq.(\ref{bigb}), $r_{h} \gg L$.

Let us also note that the electrically charged extremal black hole is dual to a zero-temperature state in the boundary field theory with a chemical potential $\mu$ given by 
\begin{equation}
\label{valmu}
\mu \sim {r_h\over L^2}.
\end{equation}
This relation is  expected from the general relation between the radial direction in gravity, $r$, and the energy scale $E$ in the field theory,
\begin{equation}
\label{enrad}
E\sim {r\over L^2}.
\end{equation}
The boundary theory lives on a sphere of radius $\sim L$, and eq.(\ref{bigb}) can be recast as 
\be
\label{recbigb}
\mu\gg {1\over L}.
\ee

The first corrections in $({r-r_h\over r_h})$ to the AdS$_2\times$ S$^2$ metric will also be important in the subsequent discussion. 
Incorporating them gives the metric,
\begin{equation}
\label{firstcorr}
\begin{split}
ds^2=  -\frac{(r-r_h)^2}{L_2^2}\bqty{1-\frac{4(r-r_h)}{3r_h}}dt^2\,&+\,\frac{L_2^2}{(r-r_h)^2}\bqty{1+\frac{4(r-r_h)}{3r_h}}dr^2\\
	&+r_h^2\bqty{1+\frac{2(r-r_h)}{r_h}}(d\theta^2\,+\,\text{sin}^2\theta \,d\varphi^2) .
\end{split}
\end{equation}
The region of the spacetime close to the horizon, including the $O({r-r_h\over r_h})$ corrections given in eq.\eqref{firstcorr}, will be called the near-horizon region below. 

Black holes close to extremality,  with a small temperature,  will also be of interest to us. The double zero in the metric coefficient $a^2(r)$ present in the extremal case will now split into two single zeros located at the outer and inner horizons  of the non-extremal   black hole.
Denoting the two horizons by 
\begin{equation}
\label{hne}
r_\pm=r_h\pm \delta r_h,
\end{equation}
where, because the black hole is still close to extremality,\footnote{For more non-extremal black holes, $r_{+} r_{-} = r_{h}^{2}$, but because of eq. \eqref{conse} we can write eq. \eqref{hne}.}
\begin{equation}
\label{conse}
{\delta r_h\over r_h}\ll 1,
\end{equation}
we have that the temperature of the black hole is 
\begin{equation}
\label{tne}
T\sim {\delta r_h\over L^2}.
\end{equation}
To avoid any confusion let us mention that in our notation  $r_h$ continues to refer to  the horizon of the extremal black hole given in terms of the charge by eq.\eqref{Qext2}, while the horizons in the non-extremal case are given by $r_\pm$, eq.(\ref{hne}).

From eq.(\ref{valmu}) and  eq.(\ref{tne}) we see that the condition eq.(\ref{conse}) is equivalent to requiring that 
\begin{equation}
\label{tmurel}
T\ll \mu,
\end{equation}
for the electrically charged case. 

Sufficiently close to the horizon, the difference between the extremal and non-extremal geometries is clearly  significant. However, once 
\begin{equation}
\label{diffr}
r-r_h \gg \delta r_h,
\end{equation}
the difference becomes small and the slightly non-extremal geometry is well approximated by the extremal one, with the same charge. Note that  condition eq.(\ref{conse}) 
 makes eq.(\ref{diffr})  consistent with eq.(\ref{condaar}). Therefore, at small temperature there is a region of the geometry  well approximated by the AdS$_2$ metric where the effect of the temperature is unimportant.
 
 In fact, since we are dealing with a big black hole, eq.(\ref{bigb}), it will be convenient in the discussion which follows to take the temperature small enough so that 
 \begin{equation}
 \label{condarh}
 {\delta r_h \ll L},
 \end{equation}
 which implies that\footnote{This corresponds to   temperatures which are smaller than the inverse-radius of the $S^2$ on which the boundary field theory lives.}
 \begin{equation}
 \label{condat}
 T\ll {1\over L}.
 \end{equation}
 This allows for a region of spacetime which lies in the asymptotic AdS$_2$ region, where eq.(\ref{csta}) is met,  and which also meets eq.(\ref{diffr}) and eq.(\ref{condaar}).
  In this region, which will play an important role in some of the discussion that follows,  the deviations due to the finite temperature have become unimportant and the metric has asymptotically attained the AdS$_2$ form, eq.(\ref{nearhor}). For brevity we will call this the asymptotic AdS$_2$ region below, see Fig.(\ref{fig:geom}). 
 Before moving ahead, we write out all the limits required to locate this region in one equation for readability,
 \begin{equation}
   \delta r_{h} \ll L_{2} \sim L \ll r-r_{h} \ll r_{h}\,.
   \label{all-lims}
 \end{equation}

 We end this subsection with some comments on thermodynamics. The excess mass, $\delta M$,  close to extremality, is related to the temperature by
 \begin{equation}
 \label{exmass}
 \delta M = \frac{\pi^2}{3G}\, T^2 L^2\,r_h\, ,
 \end{equation}
 leading to a linear specific heat,
 \begin{equation}
 \label{sph}
 C={d\delta M\over dT}= \frac{2\pi^2}{3G}\, T L^2\,r_h.
 \end{equation}
 
 We also note that the system has a gap, corresponding to a temperature very close to extremality, \cite{Shapere:1991ta,Maldacena:1996ds},
 \begin{equation}
 \label{gapex}
 T_{gap}\sim { G \over L^2 \,r_h}.
 \end{equation}
 This is much smaller than $ 1/L$. We will be working at temperatures $T\gg T_{gap}$, meeting eq.(\ref{condat}) and eq.(\ref{tmurel}).

\subsection{Time Dependence}
We now turn to studying   time-dependent perturbations of this system which preserve the spherical symmetry. 
With no loss of generality, the metric is given by
\begin{equation}
\label{metspa}
ds^2=g_{\alpha \beta}dx^\alpha dx^\beta + e^{2 \phi} d\Omega_2^2,
\end{equation}
where $g_{\alpha \beta}$ is the metric in the $r-t$ plane, and $e^{2\phi}$ is the radius of the $S^2$. These components of the metric are general functions of $(r,t)$. 
It is easy to see that the equations of motion and Bianchi identities completely determine  the gauge field which continues to be given by eqs.\eqref{Qe} and \eqref{Qm} (we use the electromagnetic duality to continue to work with the purely magnetic case for simplicity). 

By a suitable coordinate transformation the metric can be brought to the form,
\begin{equation}
\label{met2}
ds^2=-g_{tt}dt^2\,+\,g_{rr}dr^2\,+\,r^2d\Omega_2^2.
\end{equation}
The $g_{rt}$ equation of motion now leads to 
\begin{equation}
\label{grt}
\partial_tg_{rr}=0\Rightarrow g_{rr}\equiv g_{rr}(r).
\end{equation}
The $g_{tt}$ equation is first order in $r$ and can be integrated to give
\begin{equation}
\label{grrsol}
g_{rr}=\Big(1-\frac{2GM}{r}+\frac{4\pi Q_m^2}{r^2}+\frac{r^2}{L^2}\Big)^{-1}.
\end{equation}
Finally the $g_{rr}$ equation after a further time-reparametrisation gives
\begin{equation}
\label{gttsol}
g_{tt}=1-\frac{2GM}{r}+\frac{4\pi Q_m^2}{r^2}+\frac{r^2}{L^2}.
\end{equation}
As a result we see that the most general  spherically symmetric solution  of the Einstein-Maxwell 
system, in the presence of a cosmological constant, is the RN black hole.

The arguments above are simply a  proof of Birkhoff's theorem. Note in particular that the static nature of the solution was not imposed to begin with and arose as a consequence of the equations of motion, once spherical symmetry was imposed. Our conclusion, that the RN black hole is the most general spherically symmetric solution,   is to be physically  expected since there are no dynamical degrees of freedom in gravity or the gauge field in the $S$-wave sector. 

It is also worth noting that  while the analysis above was carried out in the four-dimensional theory, we would have reached the same conclusions in a two-dimensional model obtained by carrying out a dimensional reduction, as is done in section \ref{dimred}. 
The additional equations in the four-dimensional theory do not need to be used above in deriving the Birkhoff's theorem. 

One additional point needs to be examined before we conclude that there are no dynamical degrees of freedom in the $S$-wave sector of this theory. The above argument shows that there are no bulk degrees of freedom, but in the presence of a boundary extra boundary degrees can arise. However, in asymptotically AdS$_4$ space the asymptotic symmetry group is $SO(3,2)$, which is just the group of exact isometries. Thus, unlike $AdS_3$  or $ AdS_2$, here  there are no extra boundary degrees of freedom.

\subsection{Thermodynamics of the AdS$_4$ RN Black Hole}
\label{thermoads4}
The thermodynamics of the AdS$_4$ RN black hole is well understood. We summarise some key points here for completeness, more details can be found in appendix \ref{fullthermo}.

The Euclidean Einstein-Maxwell action is given by,
\begin{align}
S=-\frac{1}{16 \pi G}\int d^4 x\,\sqrt{g}\pqty{R-2\Lambda}&-\frac{1}{8\pi G}\int_{bdy} d^3 x\,\sqrt{\gamma}\,K\,+\,\frac{1}{4G}\int d^4 x \,\sqrt{g}\,F^2,\label{ads4rnact}
\end{align}
where we have included the extrinsic curvature term at the asymptotic AdS$_4$ boundary. It will be convenient in the discussion below to work with magnetically charged black holes  for which $F_{\mu\nu}$ is given in eq.\eqref{Qm}.
 The above action is divergent and we have to add counter terms to make the full 4D action finite, see \cite{Henningson:1998ey,Balasubramanian:1999re,Myers:1999psa,Emparan:1999pm,Mann:1999bt,Kraus:1999di,deHaro:2000vlm},
\begin{equation}
\label{4dcount}
S_{count}=\frac{1}{4\pi G L}\int_{bdy} d^3 x\, \sqrt{\gamma}\,\pqty{1+\frac{L^2}{4}R_3},
\end{equation}
where $R_3$ is the Ricci scalar of the boundary surface. We denote the regulated action by $S_{reg} = S+S_{count}$.

Standard manipulations then show that the entropy of the black hole is given by 
\begin{equation}
\label{ads4rnent}
S_{ent}=\beta M-S_{reg}=\pqty{\frac M 2+\frac{r_+^3}{2GL^2}-\frac{2\pi Q^2}{G}\,\frac{1}{r_+}}\beta=\frac{\pi r_+^2}{G},
\end{equation}
where $r_+$ is the outer horizon of the black hole, and the temperature is given by 
\be
\label{ads4rnT}
T=\frac{1}{2\pi}\pqty{\frac{GM}{r_+^2}-\frac{4\pi Q^2}{r_+^3}+\frac{r_+}{L^2}}.
\ee

Close to extremality, where the condition eq.(\ref{conse}) is met, the resulting  free energy is given by,
\begin{align}
\beta F&=\beta\, M-S_{ent}=\beta\,(M_{ext}+\delta M)-\frac{\pi (r_h+\delta r_h)^2}{G}\nonumber\\
&=\beta M_{ext}-\beta \,\delta M-\frac{\pi r_h^2}{G} \label{free4d}.
\end{align}
In the last line, we expanded to linear order in $\delta r_h$ and used eq.\eqref{exmass} and eq.\eqref{temper} to relate $\delta r_h$ to $\delta M$.

\section{The Four-Point Function}\label{4ptfn}
Next we couple the system to a  scalar field. The bulk scalar is dual to a scalar operator in the boundary theory which lives on $S^2\times T$ and we will be interested in the four-point function of this operator at low-energies and zero temperature.
In this paper we will restrict ourselves to working in the $S$-wave sector, invariant under the rotations of the $S^2$. We work with the metric in Euclidean signature in this section. 

The scalar $\sigma$ with mass $m$ has the  action
\begin{equation}
\label{actscalar}
S=\half \, \int d^4 x \sqrt{g} \, \left[(\partial \sigma)^2 +m^2 \si^2\right].
\end{equation}
We will consider the case $m^2>0$ below.
The scalar field is free except for gravitational interactions. 

In the S-wave sector, the system of gravity and the gauge field have no dynamical degrees of freedom. The presence of the scalar gives rise to the dynamics. 
The gravitational interactions will give rise to a non-trivial four-point function for the scalar field.
 This is analogous to effects due to the non-trivial  Coulomb field for  a  spherically symmetric charged distribution. We will be interested in the resulting four-point function.

We can expand $\sigma$ in terms of modes of definite frequency, using spherical symmetry to write it only as a function of $t$ and $r$,
\begin{equation}
\label{modeexp}
\sigma(t,r)=\int d\omega \, e^{i\omega t} \, \sigma(\omega,r),
\end{equation}
where $\sigma(\omega,r)$ in the background eq.\eqref{RNAdS4II} satisfies the equation
\begin{equation}
\label{preomf}
\frac{1}{r^2}\del_r\left(r^2a^2\del_r\sigma\right)-\Big(\frac{\omega^2}{a^2}+m^2\Big)\sigma=0.
\end{equation}

As is well-known, in the asymptotically AdS$_4$ region, $r\rightarrow \infty$,  the  $\omega$ dependent term in eq.(\ref{preomf}) can be neglected and the solution goes like
\begin{equation}
\label{asads4}
\sigma \sim  r^{\Delta_\pm},
\end{equation}
where 
\begin{equation}
\label{defdel}
\Delta_{\pm}={-3 \pm \sqrt{9+ 4 m^2 L^2} \over 2}.
\end{equation}
The $r^{\Delta_+}$ mode is the non-normalisable mode which dominates when $r\rightarrow \infty$. 
We take the  boundary of AdS$_4$ to be located at 
\begin{equation}
\label{bndads4}
{r\over L^{2}} = {1\over \delta} \gg 1,
\end{equation}
 and take the asymptotic behaviour of $\sigma$  to be 
\begin{equation}
\label{ass2}
\sigma \rightarrow \sigma(\omega) \left({r\over L^{2}}\right)^{ \Delta_+}.
\end{equation}
The coefficient $\sigma(\omega)$ is the source
in the dual field theory for frequency $\omega$. 

The on-shell action is a functional of this source term, and the four-point function in the boundary theory is given by the term in the  on-shell action  which contains four powers of $\sigma(\omega)$. 
Here, we will be interested in probing the near-extremal geometries by calculating this four-point function at sufficiently low frequencies. We will see below that   the non-trivial part of the four-point function arises 
from the near-horizon region of the spacetime in this limit.

 Let us make the required condition for the frequency to be small precise. As was mentioned above, the extremal black hole corresponds to the field theory at chemical potential $\mu$, eq.\eqref{valmu}.  We expect that the condition for small frequency  should require that 
 \begin{equation}
 \label{smomega}
 \omega \ll \mu \sim {r_h\over L^2}.
 \end{equation}
  
  In section \ref{RNBH}, after eq.(\ref{csta}) we discussed the asymptotic AdS$_2$ region which meets the conditions\footnote{Since we will be considering small temperatures,
  eq.(\ref{condarh}), eq.(\ref{csta}) ensures that eq.(\ref{diffr}) is met.} eq.(\ref{all-lims}).
  We will be interested in situations where the response to  the scalar arises   essentially  from the region extending  from the horizon to values of $r$ lying in this asymptotic AdS$_2$ region.  As we see below this will happen if the frequency dependent term in eq.(\ref{preomf}) is small for the part of spacetime lying beyond the asymptotic AdS$_2$ region; that is, at larger values of $r$.   
  
More precisely,  we see from eq.(\ref{preomf}) that the frequency dependence  can be neglected  compared to the mass term when 
 \begin{equation}
 \label{condaff}
 {\omega \over a} \ll m.
 \end{equation}
  For this to hold  in  the asymptotic AdS$_2$ region we need\footnote{We will work with non-zero mass here. A similar analysis for the massless case can also be carried out.}
 \begin{equation}
 \label{condbf}
{\omega\over m} \ll  {(r-r_h)\over L_2} .
 \end{equation}
 For $m L\sim O(1)$, we see that for $\omega$ meeting eq.(\ref{smomega}) this condition can be met for values of  $r$ lying  in the asymptotic AdS$_2$ region meeting the condition
  \be
 \label{asca}
 \omega\ll {r-r_h\over L_2^2} \ll {r_h \over L_2^2}.
 \ee

 Since  $a^2$, which is  the $g_{tt}$ component of the metric,  monotonically increases away from the horizon,  the condition eq.(\ref{condaff}) will then continue to be met by 
  increasing $r$ further,  all the way to the AdS$_4$ boundary,  and the frequency dependence in eq.(\ref{preomf})  can be neglected in this whole region away from the near-horizon spacetime.

  Once  the frequency term can be neglected eq.(\ref{preomf})  takes the form 
 \begin{equation}
\label{preomf2}
\frac{1}{r^2}\del_r\left(r^2a^2\del_r\sigma\right)-m^2\sigma=0.
\end{equation}
We see that the resulting solution, with the boundary condition eq.(\ref{ass2}) is then, to the leading order, independent of $\omega$ in this region upto a multiplicative constant $\sigma(\omega)$.  Denoting the $r$-dependent part of the solution by $f(r)$, the factorised form of the solution away from the horizon is
  \begin{equation}
  \label{siga}
  \sigma(t,r)=\int d \omega \, e^{i \omega t}  \, \sigma(\omega) \, f(r)= \sigma(t) f(r),
  \end{equation}
  where 
  \begin{equation}
  \label{defsigo}
  \sigma(t)=\int d\omega \, e^{i \omega t} \, \sigma(\omega).
  \end{equation}
  
  In contrast, in the  region sufficiently close to the horizon where eq.(\ref{condaff}) is not met and the $\omega$ dependent terms cannot be neglected in eq.(\ref{preomf}),  $\omega$ enters in the radial dependence  non-trivially. 
  
  We will see below that   in the region where this factorised form eq.(\ref{siga}) is valid,  the contribution to the four-point function is  only a 
  contact term. The non-contact terms in the time-dependence arise  solely from the region sufficiently close to the horizon where the  frequency dependence is more non-trivial.

 \subsection{On-shell Action}
 
 We now turn to computing the on-shell action which arises due to the gravitational back-reaction produced by a scalar perturbation satisfying eq.(\ref{preomf}). 
 The basic idea of the calculation is straightforward. The scalar perturbation gives rise to a stress tensor which  perturbs the metric and results in a non-trivial on-shell action.  In contrast, the gauge field  is left unchanged and does not play a role.

 The stress tensor produced by the scalar is 
 \begin{equation}
 \label{stress}
 T_{\mu\nu}=\partial_\mu \sigma \partial_\nu \sigma -\half \, g_{\mu\nu} \left[(\partial \sigma)^2 + m^2 \sigma^2\right],
 \end{equation}
 and is quadratic in the scalar.
 We are in particular interested in the four-point function on the boundary for the operator dual to the scalar. This is obtained from  terms in the on-shell action which are quartic in the source
 $\sigma(t)$, eq.(\ref{ass2}). Such  terms   arise from expressions  which are  quadratic in the stress tensor. As mentioned above we will consider S-wave perturbations for the scalar. 
 
 Our calculation is modeled along the lines of the discussion in \cite{Liu:1998ty}. 
 We expand the Euclidean metric as 
 \begin{equation}
\label{swpert}
ds^2=a^2(r)\,(1+h_{tt})\,dt^2\,+\,\frac{1}{a^2(r)}(1+h_{rr})\,dr^2\,+\,2h_{tr}\,dt\,dr\,+\,b^2(r)\,(1+h_{\theta \theta})\,(d\theta^2\,+\,\text{sin}^2\theta\,d\varphi^2),
\end{equation}
 where the perturbations are only functions of $t$ and $r$. Note that the metric perturbations which arise  also preserve spherical symmetry. 
 
 We will work in the gauge where $h_{rr}=h_{tr}=0$.  It is easy to see in general that the on-shell action  dependent on the scalar is given by 
\begin{equation}
\label{osaa}
S_{OS}=-\pi\int dt\, dr\,\pqty{\frac{b^2}{a^2}h_{tt}T_{tt}+2h_{\theta\theta}T_{\theta\theta}}.
\end{equation}
See appendix \ref{genonact} for more details. 
 
 As discussed in appendix \ref{genonact}, by using the equations of motion for the metric perturbations and the conservation equations for the stress tensor, one gets that this on-shell action is given by 
 \begin{equation}
 \label{isa}
S_{OS}=-8\pi^2 G\int dt\, dr\, \left(\frac{2a^2b^3}{b'}\,T_{rr}\frac{1}{\del_t}T_{tr}-a^2b^2\Big(1+\frac{2a'b}{b'a}\Big)\,T_{tr}\frac{1}{\del_t^2}T_{tr}\right).
\end{equation}
The $r$ integral is from the horizon to the AdS$_4$ boundary.
 We will be considering scalar perturbations with non-zero frequency here, so  the factors of inverse powers of $\partial_t$  in eq.(\ref{isa}) are  well defined and should not cause any alarm. For example, 
 ${1\over \partial_t}T_{tr} \sim {1\over ( i \omega)} T_{tr}$ if  $T_{tr} \sim e^{i \omega t}$.  
 
 We can now analyse the contribution to the $r$   integral  in eq.(\ref{isa}) from the region discussed previously where the $\omega$ dependent term in the scalar equation, eq.(\ref{preomf}), is negligible. As discussed above, in this region the scalar field takes the form eq.(\ref{siga}). 
 Inserting this in eq.(\ref{isa}) we see that the contribution to the four-point function from this region is a contact term. This is because  both terms on the RHS of eq.(\ref{isa}) involve one factor  of the stress tensor with no inverse derivative of time acting on it.  In the first term on the RHS of eq.(\ref{isa}) this  is the first factor of $T_{rr}$, and in the second term it is the factor of $T_{tr}$.  These factors of the stress tensor in turn  involve two powers of the scalar source term at the same time. As a result the contribution to the scalar four-point function is a contact term,  as is explained in more detail in appendix \ref{genonact}.
 
 We will not be interested in the contact term contributions to the four-point function here. For our purposes therefore the    four-point function can be calculated  entirely from the region where the frequency dependent term in the scalar equation cannot be neglected, and  eq.(\ref{condaff}) is not met. 
  For small frequency meeting condition eq.(\ref{condbf}), this region lies in  the near-horizon spacetime, extending from the horizon to the asymptotic AdS$_2$ region where   both eq.(\ref{condaar}) and eq.(\ref{csta}) are met.
    
\begin{figure} 
	\centering
	\includegraphics[scale=0.5]{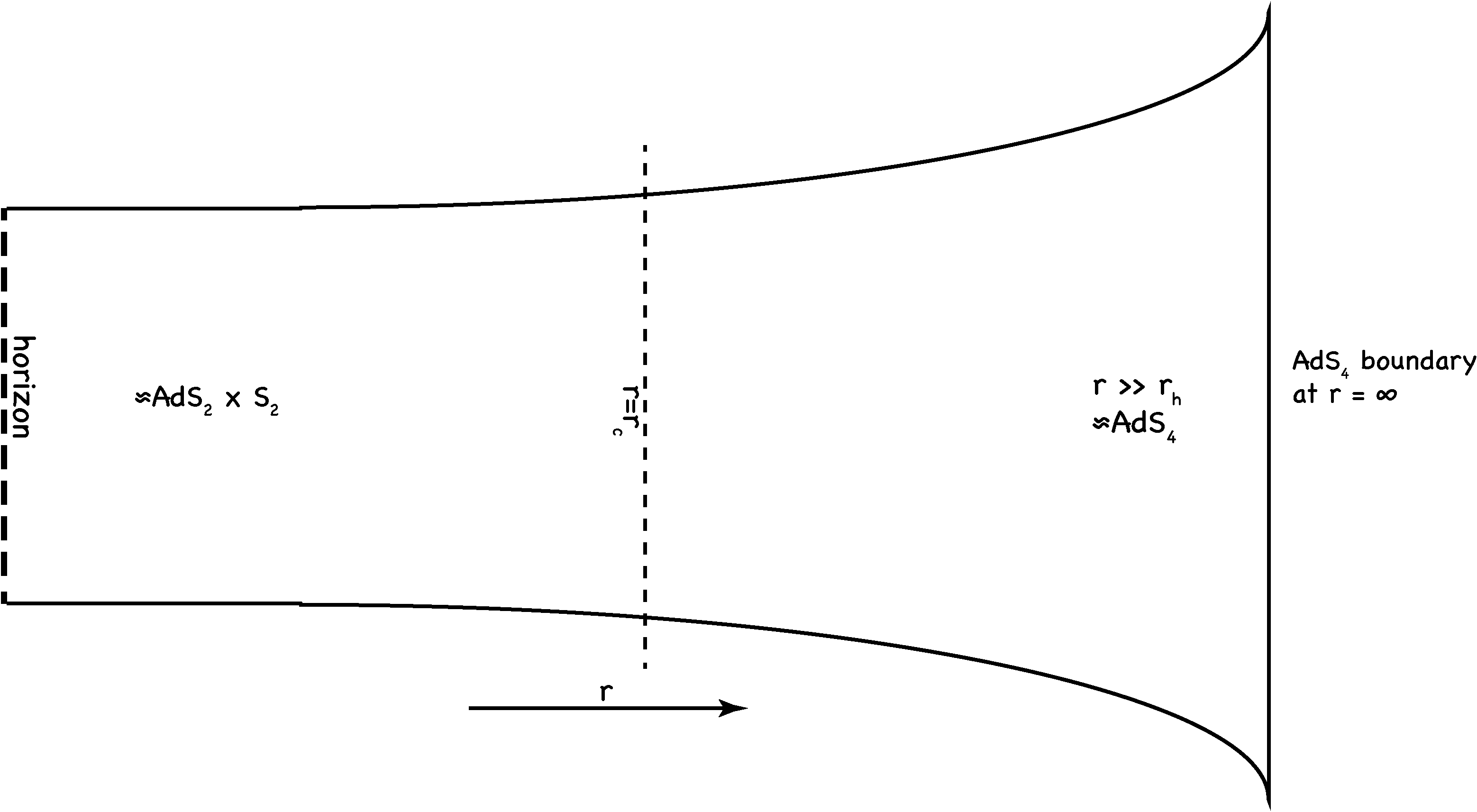}
	\caption{The near-extremal geometry. For $r\rightarrow \infty$ the geometry is asymptotically AdS$_4$. $r\rightarrow r_c$, where $\frac{r_c-r_h}{r_h}\ll 1$, $\frac{r_c-r_h}{L}\gg 1$, is the asymptotic AdS$_2\times$S$^2$ region. The horizon at extremality is at $r=r_h$.}
	\label{fig:geom}
\end{figure}

  The summary so far of this subsection then is that  we can calculate the four-point function for small frequencies by cutting off the radial integral in eq.(\ref{isa}) at a location $r=r_c$ where eq.(\ref{condaar}) and eq.(\ref{csta}) are both met,
  \begin{eqnarray}
    \text{Near-horizon limit: } & {r_c-r_h \over r_h} & \ll  1 \label{condrc1},\\
  \text{Near AdS$_2$ boundary: } & {r_c -r_h\over L_2} & \gg  1 \label{condrc2},
  \end{eqnarray} 
  and eq.(\ref{asca}) is also met so that 
  \be
  \label{condrc3}
  \omega\ll {r_c-r_h\over L_2^2}.
  \ee
  The region $r>r_c$ only gives rise to contact terms. See Fig.(\ref{fig:geom}).
  
  This result is of course what one would have expected from the relation between the energy scale in the boundary and the radial direction in the bulk. 
  Since we are interested in low frequencies, only the deep interior region of the geometry, at small values of $r$,  should have contributed as we find above. 
  We see that for sufficiently small values of $\omega$ the holographic screen can in effect be moved from the AdS$_4$ boundary to $r=r_c$ located at the boundary of the AdS$_2$ region. This process of moving the screen is the holographic analogue of moving the  RG scale in the field theory from the deep UV close to a cut-off which is closer to  the energy scales of interest.  
  
  One additional point remains to be clarified. To fully specify the calculation in terms of data only in the region $r<r_c$ we also need to relate the value of the scalar field at the AdS$_4$ boundary to its value at the new screen $r=r_c$. The value at $r_c$  effectively gives rise   to  a source in the low-energy  theory in the AdS$_2$ region. 
  As we discuss in some detail in appendix \ref{4dto2d},  up to terms which are suppressed at low frequencies, the source we introduce at the screen $r=r_c$  is related to the source in eq.(\ref{ass2}) at the AdS$_4$ boundary by a rescaling. This rescaling  is the analogue of  wave function renormalization and is to be  expected from  the analogy to the  RG transformation  in the field theory. Once this rescaling is carried out the result for the four-point function will not depend on where precisely the screen is located, i.e. on how $r_c$ is chosen.

 In view of the above discussion, the integral in the on-shell action eq.\eqref{isa} can be restricted to the near-horizon AdS$_2$ region, from the horizon to $r_c$, with the remaining integral $r>r_c$ giving rise to contact terms, 
 \begin{equation}
 \label{onshswact-restricted}
 S=-8\pi^2 G\int dt\, \int_{r_{h}}^{r_{c}} dr\, \left(\frac{2a^2b^3}{b'}\,T_{rr}\frac{1}{\del_t}T_{tr}\,-\,a^2b^2\Big(1+\frac{2a'b}{b'a}\Big)\,T_{tr}\frac{1}{\del_t^2}T_{tr}\right) + \text{contact terms}
 \end{equation}
 
 Now notice that in the second term above we can approximate  $(1+\frac{2a'b}{b'a})\simeq  \frac{2a'b}{b'a}$. For example, near $r=r_c$, which lies  in the asymptotic AdS$_2$ region, eq.(\ref{all-lims}),
 \begin{align}
 a^{2} &\approx \frac{(r-r_{h})^{2}}{L_{2}^{2}}, \nonumber\\
 a' &= \frac{1}{L_{2}}, \nonumber\\
 b &\approx r_{h}, \nonumber\\
 b' &\approx 1,
 \label{eqn:ab-nh}
 \end{align}
 so that 
 \be
 \label{rat22}
 {a'b\over b'a}\simeq {r_h\over r-r_h} \gg 1.
 \ee
 For smaller values of $r$ near the horizon the LHS in  eq.(\ref{rat22}) is even bigger since $a\rightarrow 0$. As a result the leading term in the on--shell action becomes \footnote{More correctly, the lower limit  of the $r$ integral in eq.(\ref{onshswact-restricted}), eq.(\ref{OS2}) is the outer horizon, which is at $r_h$ only in the extremal, or zero temperature, case.}
 \begin{equation}
 \label{OS2}
 S\simeq -16\pi^2 G \int dt\, \int_{r_{h}}^{r_{c}} dr\,\left(\frac{b^3a^2}{b'}\,T_{rr}\frac{1}{\del_t}T_{tr}\,-\,b^3\,a^2\frac{a'}{b'a}\,T_{tr}\frac{1}{\del_t^2}T_{tr}\right),
 \end{equation}
 where the quantities $a,b$ etc. take the values in \eqref{eqn:ab-nh}; because of the $b^{3}$, eq.\eqref{OS2} scales like $r_h^3$. 

 We will see below that this leading term is reproduced in the JT theory at zero temperature.

  In what follows, it will be convenient to use a new coordinate $z$ defined as
  \begin{equation}
  \label{zdef}
  z=\frac{L_2^2}{(r-r_h)}.
  \end{equation}
  For the extremal case we get, after converting between $r$ and $z$ coordinates,
 \begin{align}
 S 
 &\simeq  16\pi^2G\,\frac{r_h^3}{L_2^2}\int dt \int_{\delta_c}^\infty dz\,z\pqty{T_{tz}\frac{1}{\del_t^2}T_{tz}-z\,T_{tz}\frac{1}{\del_t}T_{zz}},
 \label{OSz}
 \end{align}
 with $\delta_c={L_2^2\over r_c-r_h}$.
 
 We see below that eq.(\ref{OSz}) will be reproduced by the JT model for  a suitable choice of parameters. 
 
 \section{The JT Model}
 \label{JT}
 In this section we discuss the JT model in more detail and show that it reproduces the behaviour of the four-dimensional near-extremal RN system discussed above at low energies. 
 
 The JT model was discussed in  \cite{JACKIW1985343,Teitelboim:1983ux,Maldacena:2016upp,Jensen:2016pah} and consists of a scalar $\phi$, called the dilaton,  coupled to gravity in two-dimensions with the action
  \begin{equation}
 \label{JTact}
 S_{JT}=-\frac{r_h^2}{4  G}\pqty{\int d^2 x \sqrt{g}\,R+2\int_{bdy}\sqrt{\gamma}\,K}\,-\,\frac{r_h^2}{2 G}\pqty{\int d^2 x  \sqrt{g}\,\phi\,(R-\Lambda_2)+2\int_{bdy}\sqrt{\gamma} \,\phi K}.
 \end{equation}
The first term is topological and gives rise to the ground state entropy for the extremal black hole. The second term is dynamical.  We will work in Euclidean space here. 

In section \ref{dimred}, we will carry out the S-wave reduction of the four-dimensional Einstein-Maxwell system and show that the low-energy dynamics is effectively described by the JT theory. The coefficients in the action, eq.(\ref{JTact}), have been chosen to agree with the S-wave reduction, with $G$ being the four-dimensional Newton's constant, and $r_h$ being the attractor value of the radius of the $S^2$, i.e., its value at the horizon of the extremal black hole. 
The two-dimensional cosmological constant $\Lambda_2$ is related to the AdS$_2$ radius $L_2$, eq.\eqref{ads2rad}, by 
\begin{equation}
\label{2dcc}
\Lambda_2=-\frac{2}{L_2^2}.
\end{equation}

In addition to the terms above we will see below that a counter term is needed on the boundary so that one gets  finite results for the thermodynamics and response. This is  of the form
\be
\label{Scounter}
S_{ct}={r_h^2\over G L_{2}}\int_{bdy}\sqrt{\gamma} \, \phi.
\ee

The full action then becomes
\be
\label{JTfact}
S_{JT}=-\,\frac{r_h^2}{2 G}\pqty{\int d^2 x \,\sqrt{g}\,\phi\,(R-\Lambda_2)+2\int_{bdy}\sqrt{\gamma} \,\phi K} + {r_h^2\over G L_{2}}\int_{bdy}\sqrt{\gamma} \, \phi,
\ee
where we have left out the topological term. 
 
 The equation of motion from varying $\phi$ tells us that the geometry must be AdS$_2$ with no corrections. 
 The non-trivial behaviour of this system therefore arises because of the presence of a boundary. As the boundary changes,  its extrinsic curvature and induced metric change, and this gives rise to a varying on-shell action.

To be more specific, take AdS$_2$ space (Euclidean) in Poincar{\'e} coordinates,
 \begin{equation}
 \label{ads2poin}
 ds^2=\frac{L_2^2}{z^2}\pqty{dt^2+dz^2},
 \end{equation}
  and consider a boundary in the asymptotic region, $z\ll L_2$.  
  We start with the boundary at $z=\delta, {\delta\over L_2}\ll 1$.
For small fluctuations, the resulting boundary can now be described by the curve, 
\be
\label{flucb}
z (1- \epsilon'(t))=\delta,
\ee
where $\epsilon(t)$ parametrises the fluctuations. 

It is convenient to change coordinates,
\begin{align}
t & =  \hat t+\epsilon(\hat t)-{\hat z^2 \epsilon''(\hat t)\over 2} \label{ct1},\\
z& = \hat z(1+\epsilon'(\hat t)) \label{ct2}.
\end{align}
In these new coordinates the metric takes the form
\be
\label{newmet}
ds^2={L_2^2 \over {\hat z^2} }\, (1+h_{tt}) \,d\hat t^2 + {L_2^2\over {\hat z^2}} \, d \hat z^2,
\ee
with 
\be
\label{valhtt}
h_{tt}=-\epsilon'''(\hat t) \hat z^2,
\ee
and the boundary is at $\hat{z}=\delta$. 
The fluctuations are now parametrised by $h_{tt}$. Eq.(\ref{newmet}) corresponds to choosing the Fefferman-Graham (FG) gauge. 

We will find it convenient to work in the $(\hat t,\hat z)$ coordinates below, in which the boundary is fixed at coordinate value $\delta$ and  $h_{tt}$ is turned on. To avoid clutter we will drop the hats on the new coordinates and refer to them now on  as $(t,z)$, with a metric of the form eq.(\ref{newmet}). 
We see that the coordinate transformation eq.(\ref{ct1}), eq.(\ref{ct2}) involves a time reparametrisation, and from eq.(\ref{newmet}) we see that this transformation  is an  asymptotic isometry, since $h_{tt}$ vanishes as $z\rightarrow 0$, but is not zero in general.

More generally, even in the presence of other fields which we will introduce later,  the boundary conditions we will impose on the field $\phi$ and $h_{tt}$ are as follows,
\begin{align}
\phi &= {\alpha\over z} \label{asdil1},\\
h_{tt} &\rightarrow-\epsilon'''(t)  z^2 \label{asmet1} ,
\end{align}
as $z\rightarrow \delta$. 
These correspond to the dilaton being fixed (Dirichlet boundary condition) and to $h_{tt}$ vanishing, with the coefficient of the leading $z^2$ behaviour being a general function of time. 

Let us now show that for the action eq.(\ref{JTfact}) these boundary conditions give  rise to a well defined variational principle. 

Expanding the action  to quadratic order we get
\be
\label{quadact1}
S=-{r_h^2\over 2G}\int d^2x \phi\left[-\partial_z^2h_{tt}+{\partial_z h_{tt}\over z}\right] + {r_h^2\over 2G}\int_{bdy} \phi \,\partial_z h_{tt}.
\ee
The boundary terms obtained when carrying out a variation of this action  take the form
\be
\label{bt}
S_{bt}={r_h^2\over 2G}\int_{bdy} \delta \phi \partial_z h_{tt} + \left(\partial_z \phi + {\phi\over z}\right) \delta h_{tt},
\ee
where $\delta \phi$ and $\delta h_{tt}$ are the variations. 
For a well defined variational principle to exist these boundary terms must vanish. We see that for the boundary conditions mentioned above,  this is indeed true.

 From eq.(\ref{quadact1}) we find that the  action, for a solution with metric being AdS$_2$ and with the behaviour eq.(\ref{asdil1}), eq.(\ref{asmet1}), is given by
 \be
 \label{Sons}
 S=-{r_h^2\over G} \, \alpha \int_{bdy} \epsilon'''(t).
 \ee
 We see that the action depends on  $\epsilon(t)$ which is the time reparametrisation degree of freedom associated with fluctuations in the boundary. The $SL(2,R)$  isometries of AdS$_2$ correspond to $\epsilon'''=0$ and give rise to a vanishing contribution. More generally, one can argue (see \cite{Kitaev-talks:2015, Maldacena:2016upp})  that an action  for $\text{diff}/SL(2,R)$,\footnote{diff denotes time reparametrisations.} which  at the linearised level goes like $\epsilon'''$, must be proportional to the Schwarzian derivative of the time reparametrisations,
 \be
 \label{Sons2}
 S=-{r_h^2\over G}\,\alpha \int_{bdy}  \text{Sch}[\epsilon(t)],
 \ee
 where Sch$[\epsilon(t)]$, the Schwarzian derivative, is defined  for a time reparametrisation as follows,
 \begin{eqnarray}
 f(t) & = & t+ \epsilon(t), \\ 
 \text{Sch}[\epsilon(t)] & =  & -{1\over 2 } {(f'')^2\over (f')^2}+ \pqty{{f''\over f'}}'.
 \end{eqnarray}
 Such a more general action would arise if we started with eq.\eqref{JTfact} and kept terms beyond the quadratic order in $\phi, h_{tt}$. 
 
 For reference below we note that the Schwarzian to quadratic order in $\epsilon(t)$ is given by
 \be
 \label{schwquad}
 \text{Sch}[\epsilon(t)] = \epsilon''' -{3\over 2}(\epsilon'')^2-\epsilon'\epsilon''' + \cdots
 \ee
 From the quadratic terms we see that the equation of motion, at the linear level for $\epsilon(t)$, is 
 \be
 \label{eomep}
 \epsilon''''=0,
 \ee
 with the most general solution
 \be
 \label{soleomep}
 \epsilon(t) = \alpha_0+\alpha_1t +\alpha_2t^2+\alpha_3t^3.
 \ee
 Here, $\alpha_0,\alpha_1,\alpha_2$ parametrise $SL(2,R)$ transformations, while $\alpha_3$ is related to the mass of the non-extremal black hole. 
 The fact that the most general solution, up to diffeomorphisms, is a black hole agrees with Birkhoff's theorem applied to the $S$-wave sector of the $4$ dimensional system.\footnote{We thank Juan Maldacena and Douglas Stanford for emphasizing this to us.}

 In summary, the dynamics of the JT model arises due to fluctuations of the boundary. These are related to time reparametrisations eq.(\ref{ct1}), eq.(\ref{ct2}), and are described by an action involving the Schwarzian derivative, eq.(\ref{Sons}), eq.(\ref{Sons2}).

 \subsection{Thermodynamics}
 A time independent solution of the action eq.\eqref{JTfact} is given by
 \be
 \label{valphi}
 \phi={L_2^2\over r_h z},
 \ee
 with the metric being eq.\eqref{newmet}. 
 
 In this solution the $SL(2,R)$ isometry of AdS$_2$ space is broken by the non-vanishing dilaton. This solution meets the boundary conditions eq.\eqref{asdil1}, eq.\eqref{asmet1},
 with $\alpha =\frac{L_2^2}{r_h}$. 
 This  solution is the starting point for understanding the low-energy behaviour of near-extremal black holes in four dimensions, as we will see below.

 Black holes in the JT model are given by the metric
 \begin{equation}
 \label{JTBH}
 ds^2=\pqty{\frac{(r-r_h)^2}{L_2^2}-\frac{2G\delta M}{r_h}}dt^2+\frac{dr^2}{\pqty{\frac{(r-r_h)^2}{L_2^2}-\frac{2G\delta M}{r_h}}},
 \end{equation}
 with the dilaton given by eq.(\ref{valphi}). Here $r$ is related to $z$ by eq.\eqref{zdef}.
  
 The resulting on-shell action is easy to calculate and gives (see appendix \ref{thermoap} for details)
 \be
 \label{FJT}
 S_{JT}=-\beta \delta M-\frac{\pi r_h^2}{G}.
 \ee
 Note that the counter term eq.(\ref{Scounter}) is needed to get a finite result. 
 
We see that this almost agrees with the free energy eq.\eqref{free4d} obtained from the four-dimensional analysis of the near-extremal geometry; there is an extra term $\beta M_{ext}$ in eq.\eqref{free4d}. This tells us that the near-horizon system does not reproduce the mass of the extremal state, but it does correctly capture the departures from extremality.\footnote{This is similar to what happens  in the near-horizon AdS$_5\times S^5$ geometry for $D3$ branes. This region does not  correctly reproduce the  ADM mass of the $D3$ branes, but it does give rise accurately to the low-energy dynamics which agrees with that of the $\mathcal{N}=4$ SYM theory.}

 It is also easy to see that the topological term, eq.\eqref{stop}, gives the correct ground state entropy. Thus the JT model correctly reproduces the thermodynamics of the near-extremal RN black hole. 
 
 \subsection{Four-Point Function}
 We now introduce an extra scalar field $\sigma(t,z)$, with the action
  \be
  \label{actsc1}
  S_{\sigma}=2 \pi r_h^2 \int d^2x\sqrt{g}\,\pqty{ (\partial\sigma)^2+ m^2 \sigma^2}.
  \ee
  Note that the scalar field only couples to the metric and not to the dilaton. 
  It is easy to see from eq.(\ref{actsc1}) that the asymptotic behaviour of $\sigma(t,z)$ as $z\rightarrow 0$ is 
  \be
  \label{assscalar}
  \sigma \rightarrow \sigma_+(t) z^{\Delta_+} + \sigma_-(t) z^{\Delta_-} ,
  \ee
  where 
  \be
  \label{defdpm}
  \Delta_\pm={1\pm \sqrt{1+4 m^2L_2^2}\over 2}.
  \ee
  The normalisable and non-normalisable modes correspond to the $z^{\Delta_+}, z^{\Delta_-}$  behaviour respectively.\footnote{We take $m^2>0$ for simplicity here.}
  
 For small fluctuations, the boundary can be described by eq.\eqref{flucb}. We impose Dirichlet boundary condition on this boundary to make the variational principle well defined. The scalar action comes out to be (see appendix \ref{secoup} for details)
 
 \begin{equation}
 \label{scalaract2}
 S_{\sigma}=4\pi r_h^2\,C\,\Delta_+\,(\Delta_--\Delta_+)\int dt_1\,dt_2\,\frac{\sigma(t_1)\,\sigma(t_2)}{|t_1-t_2|^{2\Delta_+}}\pqty{\epsilon'(t_1)-2\,\frac{\epsilon(t_1)}{|t_1-t_2|}}.
 \end{equation}
 Expressing the action eq.\eqref{scalaract2} in terms of the stress tensor components,
 \be
 \label{defT}
 T_{\alpha\beta}=\partial_\alpha \sigma \partial_\beta \sigma -{1\over 2}\, g_{\alpha \beta}\left((\del\sigma)^2+m^2 \sigma^2\right),
 \ee
 we get
 \begin{equation}
 \label{scalaract3}
 S_{\sigma}=4\pi r_h^2\int dt \,\pqty{\epsilon'(t)zT_{zz}+\epsilon(t)T_{tz}}.
 \end{equation}
 
 We can now compute the on-shell action for the JT model.
 The total action is given by
 \begin{equation}
 \label{totact2}
 S=S_{\epsilon}+S_{\sigma},
 \end{equation}
 where $S_{\epsilon}$ is the Schwarzian action defined in eq.\eqref{Sons2}.
 We work to quadratic order in $\epsilon(t)$,  eq.(\ref{schwquad}).  This gives
 \be
 \label{onsact2}
 S_{\epsilon}=-{r_hL_2^2\over G}\int dt \pqty{-{3\over 2 }(\epsilon'')^2-\epsilon'\epsilon'''}.
 \ee
 Here we have substituted for $\alpha$ from eq.(\ref{valphi}). 
 Carrying out integration by parts and dropping total derivatives we then get 
 \be
 \label{onsa3}
 S_{\epsilon}={r_hL_2^2\over 2 G}\int dt \,\epsilon(t)\, \epsilon''''(t).
 \ee
 Therefore the total action eq.\eqref{totact2} is
 \begin{equation}
 \label{totact3}
 S={r_hL_2^2\over 2 G}\int dt \,\epsilon(t)\, \epsilon''''(t)\,+\,4\pi r_h^2\int dt \,\pqty{\epsilon'(t)zT_{zz}+\epsilon(t)T_{tz}}.
 \end{equation}
 We find the equation of motion for $\epsilon(t)$ from eq.\eqref{totact3} to be
 \be
 \label{4de}
 \epsilon''''(t)=-{4\pi G r_h\over L_2^2}(T_{tz}-z\partial_t T_{zz}).
 \ee
 As a result, the on-shell action becomes
 \begin{align}
 S_{OS}&={2 \pi r_h^2} \int dt \, \epsilon\, (T_{tz}-z \partial_t T_{zz})\nonumber\\
 &=-\frac{8\pi^2 G r_h^3}{L_2^2}\int dt\, (T_{tz}-z \partial_t T_{zz})\,\frac{1}{\del_t^4}\,(T_{tz}-z \partial_t T_{zz}). \label{onsa4}
 \end{align}
  (We are considering time dependent solutions for $\sigma$ so the factors of $1\over \partial_t$ should not cause any alarm, as in section \ref{4ptfn} above).
 Note that the action eq.\eqref{onsa4} is expressed as a time integral on the boundary. We can express it as a bulk integral as follows,
 \begin{align}
 S_{OS}&=\frac{8\pi^2 G r_h^3}{L_2^2}\int dtdz\,\del_z\pqty{ (T_{tz}-z \partial_t T_{zz})\,\frac{1}{\del_t^4}\,(T_{tz}-z \partial_t T_{zz})}\nonumber\\
 &=\frac{16\pi^2 G r_h^3}{L_2^2}\int dtdz\,\del_z\pqty{ T_{tz}-z \partial_t T_{zz}}\,\frac{1}{\del_t^4}\,(T_{tz}-z \partial_t T_{zz})\label{onsa5}.
 \end{align}
 Note that the extra minus sign is due to the boundary, $z=\delta$, being evaluated at the lower limit of the $z$ integral.
 Note also that a possible contribution from the horizon vanishes because the stress tensor vanishes there. 
 
 We can use the stress tensor conservation equations, eq.\eqref{stresscons} in appendix \ref{dimredsource}, to simplify eq.\eqref{onsa5} to get
 \begin{equation}
 \label{onsa6}
 S_{OS}=\frac{16\pi^2 G r_h^3}{L_2^2}\int d^2x\,z\,T_{tz}\,\frac{1}{\del_t^2}\,(T_{tz}-z \partial_t T_{zz}).
 \end{equation}
 
 Note that this agrees with the action eq.\eqref{OSz}.
 Thus, we see that the JT model agrees with the result obtained from the four-dimensional analysis at low-energies. 
 
 Let us next turn to the solution for the dilaton in the presence of the scalar $\sigma$. 
 As discussed in appendix \ref{dimredsource} the  solution for $\phi$ is
 \begin{equation}
 \phi=\frac{L_2^2}{r_h z}\pqty{1-\epsilon'(t)}+\frac{4\pi G}{z}\pqty{z\frac{1}{\del_t^2}T_{zz}-\frac{1}{\del_t^3}T_{tz}}\label{phisol2}.
 \end{equation}
 From eq.\eqref{4de} and eq.\eqref{phisol2}, we see that the dilaton is a constant on the boundary.
 
 It is interesting to note that the value of $\epsilon(t)$ obtained in eq.(\ref{4de}) by solving the Schwarzian action coupled to the scalar source automatically yields a solution which makes the dilaton constant on the boundary, consistent with the boundary conditions we imposed above. 
 
 Note also that the prefactors in eq.(\ref{onsa5}), eq.(\ref{onsa6}) go like $r_h^3$, and are  enhanced by an extra factor of $r_h$ compared to the coefficient of the JT action we started with, eq.(\ref{JTact}), and also the coefficient of the scalar action, eq.(\ref{actsc1}).  These two coefficients, which are chosen to agree with what we  obtain after carrying out the dimensional reduction as explained in the next section, go like ${r_h^2\over G}$, with $G$ being the four-dimensional Newton's constant. This is because   the dimensional reduction is carried out over an    $S^2$ of radius $r_h$.\footnote{For the scalar action eq.\eqref{actsc1} we have absorbed the factor of $1/G$ by rescaling $\sigma$.}

 The reason for the enhancement  of the prefactor in eq.(\ref{onsa5}), eq.(\ref{onsa6}) is tied to the fact that the Schwarzian action in eq.\eqref{totact3}  has a coefficient of order $r_h$,
 which is therefore suppressed by a factor of $1/r_h$ compared to the coefficients in eq.(\ref{JTact}), eq.(\ref{actsc1}). The Schwarzian action arises because the time reparametrisations are broken; this in turn is tied to the breaking of the $SL(2,R)$ symmetry of AdS$_2$ due to the running of the dilaton.  Since this running of the dilaton is suppressed by a factor of $1/r_h$, eq.(\ref{valphi}),  the resulting coefficient of the Schwarzian is also suppressed.  This suppression then results in the on-shell value for $\epsilon(t)$ being big and going like $r_h$, eq.\eqref{4de}, and in turn results in the on-shell action, eq.(\ref{onsa5}), eq.(\ref{onsa6}), being $O(r_h^3)$. 
 
 These features are entirely  analogous to what happens in the SYK model, with the parameter $r_h/L^2$ playing the role of $J$ - the coefficient of the four Fermi coupling. In the electrically charged case $r_h\over L^2$ is also of order the chemical potential $\mu$ in the boundary theory, eq.\eqref{valmu}. It is interesting that $r_h^2/G$ also determines the ground state entropy of the extremal system.

 In summary,  we have seen above that in the JT model the scalar source back-reacts on the dilaton and pushes the boundary in or out, causing it to fluctuate. The resulting on-shell action then gives rise to the four-point function. We have worked with the Euclidean theory here; by suitably analytically continuing in the standard fashion, one can obtain Minkowski correlators, including the out of time order four-point correlator, \cite{Roberts:2014ifa,Shenker:2014cwa,Maldacena:2016upp}. These will  continue to agree between the JT model discussed here and the low-energy limit of the four-dimensional theory in section \ref{dimred}.

\section{S-wave Reduction}
 \label{dimred}
 In this paper we have been studying spherically symmetric configurations in four dimensions. 
 Here we will explicitly construct the two-dimensional model obtained by carrying out the dimensional reduction of the four-dimensional theory. 
 The resulting two-dimensional model has some interesting differences with the Jackiw-Teitelboim theory which we will comment on below. 
 On coupling the system to a scalar field we will find that the resulting four-point function, which receives its contribution at low-energies from the near-horizon AdS$_2$ region,
 agrees to leading order in the approximations with the result eq.(\ref{OSz}) obtained above. Moreover, our analysis in the two-dimensional theory will reveal that this leading order result arises because of the dynamics associated with the boundary of the near-horizon AdS$_2$ region, which is described by an action involving the Schwarzian derivative of time reparametrisations. 
 In this section we work in Euclidean space.

 The two-dimensional model we will consider is obtained by starting with  the four-dimensional Euclidean action
 \begin{equation}
 \label{actem}
 S = -\,\frac{1}{16\pi G} \int d^4x \sqrt{\hat{g}} \, \Big(\hat{R} - 2 \hat{\Lambda}\Big)- \, \frac{1}{8\pi G} \int d^3x \, \sqrt{\hat{\gamma}} \, K^{(3)} + \frac{1}{4 G}  \int d^4x \sqrt{\hat{g}} \, F^2,
 \end{equation}
 where $\hat\gamma$ denotes the determinant of the induced metric on the three-dimensional boundary, and $K^{(3)}$ is the trace of the extrinsic curvature of the boundary. We have used a hat, $\hat{\mbox{}}$ , to denote four-dimensional quantities. We will be considering the case where we have a magnetic charge, see eq.\eqref{Qm}. Note that we have included the Gibbons-Hawking boundary term in the action above. 
 
 For dimensional reduction, we assume the four-dimensional metric to have the form
 \begin{equation}
 \label{le4d}
 ds^2 = g_{\alpha\beta}(t,r) \, dx^\alpha dx^\beta + \Phi^2(t,r) \, d\Omega_2^{\,2},
 \end{equation}
 where $g_{\alpha\beta}$ is the two-dimensional part of the metric and the dilaton $\Phi$, which is the radius of the $2$-sphere,  is assumed to be independent of the angular coordinates $(\theta, \varphi)$. 
 
 We take a magnetically charged black hole and solve for the magnetic field in terms of the metric, eq.\eqref{Qm}.  
 
 Substituting for the metric from  eq.\eqref{le4d} and for the magnetic field in the action eq.\eqref{actem}, and carrying out the integrals over the $(\theta, \varphi)$ coordinates gives the two-dimensional action
 \begin{equation}
 \begin{split}
 \label{actms}
 S = - &\frac{1}{4G} \int d^2x \sqrt{g} \, \Big[ 2 + \Phi^2 (R - 2 \hat{\Lambda}) + 2 (\nabla\Phi)^2\Big]
 +\frac{2\pi Q_m^2}{G} \int d^2x \sqrt{g} \, \frac{1}{\Phi^2}\\
 &+ \frac{1}{G} \int d^2x \, \del_\mu \big(\sqrt{g} g^{\mu\nu} \Phi \del_\nu\Phi \big) - \frac 1 G\int_{bdy} \sqrt{\gamma}\,\Phi \,n^{\alpha}\del_{\alpha}\Phi -\frac{1}{2G}\int_{bdy}\sqrt{\gamma}\,\Phi^2 K.
 \end{split}
 \end{equation}
 Here $n^{\alpha}$ is the outward pointing unit normal vector to the boundary. The first two terms on the second  line, which are quadratic in $\Phi$,
 cancel. This gives the action
 \begin{equation}
 \begin{split}
 \label{actms}
 S = - &\frac{1}{4G} \int d^2x \sqrt{g} \, \Big[ 2 + \Phi^2 (R - 2 \hat{\Lambda}) + 2 (\nabla\Phi)^2\Big]
 +\frac{2\pi Q_m^2}{G} \int d^2x \sqrt{g} \, \frac{1}{\Phi^2}\\
 &  -\frac{1}{2G}\int_{bdy}\sqrt{\gamma}\,\Phi^2 K.
 \end{split}
 \end{equation}
 
 The boundary we consider will be located in the asymptotically AdS$_2$ region described in section \ref{RNBH}, see also section \ref{4ptfn}. This is  a region where the $r$ coordinate takes a  value meeting both eq.(\ref{csta})  and eq.(\ref{condaar}). We will take the boundary to be at $r=r_c$, as  in section \ref{4ptfn} eqs.(\ref{condrc1}), (\ref{condrc2}) and eq.(\ref{condrc3}); also see Fig.(\ref{fig:geom}). 
 
 Eq.\eqref{actms} is our two-dimensional theory. It is straightforward to verify that AdS$_2$ is a solution with the dilaton taking a constant value, $r_h$. In fact this is the AdS$_2\times$S$^2$ solution of the four-dimensional theory, eq.\eqref{nearhor}.

 \subsection{Comparison with the JT Model}
 
 In order to make contact with the JT model, eq.\eqref{JTact}, let us go to a frame where the dilaton kinetic energy  vanishes. This can be achieved by performing a Weyl rescaling of the two-dimensional metric
 \begin{equation}
 \label{weyl}
 g_{\alpha\beta} \rightarrow \frac{r_h}{\Phi} \, g_{\alpha\beta},
 \end{equation}
 with $r_h$ being the  value of the dilaton in the AdS$_2 \times$S$^2$ solution described above. The action eq.\eqref{actms} now becomes
 \begin{equation}
 \begin{split}
 \label{actweyl}
 S = &- \frac{1}{4G} \int d^2x \sqrt{g} \, \bigg[ \frac{2r_h}{\Phi} + \Phi^2 R - 2 r_h  \Phi \hat{\Lambda}\bigg] +\frac{2\pi Q_m^2}{G} \int d^2x \sqrt{g} \, \frac{r_h}{\Phi^3}\\
 &-\frac{1}{2G}\int_{bdy}\sqrt{\gamma}\,\Phi^2 K.
 \end{split}
 \end{equation}

 We now expand the dilaton in terms of a perturbation about its value $r_h$ in the AdS$_2\times$S$^2$ solution described above, 
 \begin{equation}
 \label{flucdil}
 \Phi=r_h(1+ \phi).
 \end{equation}
 Inserting this in the action eq.(\ref{actweyl})  and expanding to quadratic order in $\phi$ gives
 \begin{equation}
 \begin{split}
 \label{qacte3}
 S = &- \frac{r_h^2}{4G}\pqty{ \int d^2x \sqrt{g}\, R +2\int_{bdy}\sqrt{\gamma}\,K} - \frac{r_h^2}{2G} \int d^2x \, \sqrt{g}\,\phi\pqty{R-\Lambda_2}
 + \frac{3 r_h^2 \,\kappa}{G\,L_2^2} \int d^2x \, \sqrt{g}\,\phi^2\\
  &-\frac{r_{h}^{2}}{G}\int_{bdy}\sqrt{\gamma}\,\phi K-\frac{r_h^2}{2G}\int_{bdy}\sqrt{\gamma}\,\phi^2 K.
 \end{split}
 \end{equation}
Here
$\kappa$ is defined as
 \begin{equation}
 \label{defkappa}
 \kappa = {L^2 + 4r_h^2 \over L^2 + 6 r_h^2 },
 \end{equation}
and takes the value $\kappa \simeq {2\over 3}$ when $r_h\gg L$.

 Comparing with eq.(\ref{JTact}) we see that the dimensionally reduced model has an extra term in the bulk, which is the third term in eq.(\ref{qacte3}), going  like $\phi^2$,
 and also an extra boundary term, which is the last term in eq.(\ref{qacte3}), going like $\phi^2 K$. 
 
 One important consequence of the extra bulk term is that the metric in the solution  for the dimensionally reduced model in the presence of  a varying dilaton is not AdS$_2$ and  has corrections, and these arise at the same order as the varying dilaton.  This is easy to see because the   equation of motion for $\phi$ in the presence of this term becomes  schematically of the form
 \begin{equation}
 \label{phieom}
 R=\Lambda_2+\mathcal{O}(\phi),
 \end{equation}
 which means that the metric must depart from AdS$_2$ at linear  order as $\phi$. 
 
 In fact this agrees with what we already know from the near-horizon metric in the four-dimensional solution, eq.(\ref{RNAdS4II}), eq.(\ref{firstcorr}).  Expanding eq.(\ref{RNAdS4II}) for the extremal case in  the near-horizon  limit, we get up to terms of order $O({r-r_h\over r_h})$,     
 \begin{equation}
 \label{phisol}
 \phi=\frac{r-r_h}{r_h}
 \end{equation}
 and
 \begin{equation}
 \label{2dmetsol}
 ds^2=\frac{(r-r_h)^2}{L_2^2}\pqty{1-\frac{(r-r_h)}{3 r_h}}dt^2\,+\,\frac{L_2^2}{(r-r_h)^2}\pqty{1+\frac{7(r-r_h)}{3 r_h}}dr^2.
 \end{equation} Here we have rescaled the  two-dimensional metric  from its value in eq.(\ref{RNAdS4II}) to agree with eq.(\ref{weyl}).
 We see from  eq.(\ref{2dmetsol}) and eq.(\ref{phisol})  that the metric also changes to the same order in $1/r_h$  as the dilaton.

 Let us end this subsection with one more comment. To completely specify the dimensionally reduced model we also need to specify the boundary conditions which must be met at the boundary located at $r=r_c$, eq.(\ref{condrc1}),  eq.(\ref{condrc2})  and eq.(\ref{condrc3}). 
It is clear from the discussion in section \ref{4ptfn} that for the probe scalar we introduce below we must impose Dirichlet boundary conditions. In addition we will also impose Dirichlet boundary conditions on the dilaton, like in the JT model, eq.(\ref{asdil1}).\footnote{
We will see below that the extra terms present in the action of this model, which are absent in the JT theory,  do not contribute to the leading order behaviour at low energies.
A striking fact about the JT model, as noted after eq.(\ref{phisol2}),  is that the  value for the dilaton obtained by solving for $\epsilon$ in eq.(\ref{4de}) automatically satisfies Dirichlet boundary conditions at the boundary. This feature also leads us to   impose Dirichlet boundary conditions for the dilaton in  the dimensionally reduced model.}

 \subsection{More Detailed Comparison with the JT Model}
  
 Despite the presence of extra terms in the action of   the dimensionally reduced model,  we will see here that close to extremality
 the thermodynamics and response to a probe scalar agrees  with the JT model.
 
 We begin with the thermodynamics. 
  We are interested in working to leading order in ${TL^2\over r_h}$ and ${L\over r_h}$. In the discussion below we only keep track of powers of $r_h$,
  the appropriate factors of $L$ can then be inserted on dimensional grounds.\footnote{Also, since $L_2\simeq {L\over \sqrt{6}}$, we do  not distinguish between these two scales for our parametric estimates here.}
  
  The partition function is obtained from the on-shell action of the black hole solution.
  It is easy to see that the two extra terms in eq.(\ref{qacte3}) compared to eq.(\ref{JTact}) make a subleading contribution to the action by themselves. 
  This is because both terms are quadratic in $\phi$. Since $\phi$ is $O(1/ r_h)$, eq.(\ref{phisol}), it follows  that these  quadratic terms are  suppressed compared to terms linear in the dilaton by $O(1 / r_h)$.
  
  In addition, as argued above, the bulk  term going like $\phi^2$ also causes a departure in the metric to leading order in $1/ r_h$. 
  However this correction to the metric   can be neglected in computing the on shell action.  The extra bulk  term causes the metric of the black hole in the 
  JT model, eq.(\ref{JTBH}), to be modified to
  \begin{align}
ds^2&=a(r)^2\,dt^2\,+\,a(r)^{-2}\,dr^2,\nonumber\\
a(r)^2&=\frac{(r-r_h)^2}{L_2^2}\bqty{1-\frac{4(r-r_h)}{3r_h}}-\frac{2G\delta M}{r_h}\bqty{1-\frac{r-r_h}{r_h}}\,+\,O\pqty{\frac{1}{r_h^2}}\label{dimredmet}.
\end{align}
We see that the corrections  due to the presence of the $\phi^2$ term are suppressed by  $O(1/r_h)$. As a result 
the  extra terms in the action are also suppressed by  $O(1/ r_h)$. 
Note that the effect of the Weyl rescaling, eq.\eqref{weyl}, in the computation of the action can be neglected for the same reason - it contributes terms which are suppressed.

 Let us now argue why the four-point function  for the probe scalar at leading order is the same as that in the JT theory. The probe scalar in the dimensionally reduced model also couples to the dilaton $\phi$, this is the last term going like $\int \phi J$ in eq.(\ref{qacte}) of appendix \ref{dimredsource}. This gives rise to one additional term in the dimensionally reduced case, besides the two terms quadratic in $\phi$ present in eq.(\ref{qacte3}).
 All  of these terms however only contribute to subleading order in $1/ r_h$. 
 
 Let us start with the bulk term quadratic in $\phi$ present in eq.(\ref{qacte3}). 
 The four-point function is obtained from the on-shell action quartic in $\sigma$. The dilaton solution sourced by $\sigma$ takes the forms
 \begin{equation}
   \phi \sim \frac{L_{2}^{2}}{r_{h} z} (1 - \epsilon'(t)) + F[\sigma].
   \label{phi-soln-schematic}
 \end{equation}
  This is discussed in eq.\eqref{phifluc} of appendix \ref{dimredsource}, with $F[\sigma]$ given by eq.\eqref{fsig}, and being of  order unity in the $1/ r_h$ expansion.
  Since $\epsilon(t)$ is $O(r_h)$, eq.(\ref{4de}),  we see   that the $\epsilon(t)$ dependent term above  is $ O(r_h^0)$. 
  Since the prefactor of the bulk term  under question in the action eq.(\ref{qacte3}) goes like $r_h^2$, it then follows that the  contribution this term makes to the on-shell action also goes like 
  $r_h^2$. In contrast, we have seen from eq.(\ref{OSz}), eq.(\ref{onsa6}) that the leading contribution goes like $r_h^3$. Thus we see that this extra bulk term makes a contribution suppressed by $O(1/ r_h)$. 
  
  The extra boundary term in eq.(\ref{qacte3}) which goes like $\int \sqrt{\gamma} \phi^2 K$ also makes a suppressed contribution. For example, taking $\phi$ to be its background value, eq.(\ref{phisol}),   gives a contribution suppressed  by a factor of $1/r_h$ compared to the leading term which goes like $\int \sqrt{\gamma}\phi K$. 
  
  Finally, we consider the additional bulk term which arises due to the $\int  \phi J$ coupling  in the bulk, eq.\eqref{qacte}. 
  As argued above,  $\phi \sim O(r_h^0)$, and $J$ is also $O(r_h^0)$.  Thus, taking  the prefactor of this term which is $O(r_h^2)$ into account, we see that the net contribution it makes is $O(r_h^2)$, which is down compared to the leading term by a factor of $1/r_h$.

  In summary, the extra terms which arise in the dimensionally reduced model can be neglected, compared to the terms present in the JT model, in calculating the  thermodynamics at low temperatures and the response to a probe scalar at low frequencies, when working  to leading order in $L/r_h$.

\section{Conclusions}
In this paper we have studied near-extremal black holes in asymptotically AdS$_4$ spacetime in the $S$-wave sector. These black holes arise in a theory consisting of gravity coupled to a Maxwell field. We analysed big black holes meeting the condition $r_h\gg L$, with $r_h,L$ being the black hole horizon radius and the AdS radius respectively, eq.(\ref{bigb}), and studied  
 the near-extremal thermodynamics and the response of the black hole  to a probe scalar field. We find that the dynamics, at low energies and to leading order in the parameter $L/r_h$,   is well approximated by the Jackiw-Teitelboim theory of gravity.\footnote{The  parameter ${r_h\over L^2} $ in this system is analogous to the coefficient $J$ of the four Fermi coupling in the SYK model, \cite{Maldacena:2016hyu}. For an electrically charged black hole ${r_h\over L^2 }$  is of order $\mu$ -  the chemical potential in the boundary theory.}  In fact,  the low-energy dynamics is determined by symmetry considerations alone, with the JT theory being the simplest realisation of these symmetries.

Our analysis shows that the low-energy dynamics arises from the near-horizon  AdS$_2$ region of the spacetime. This region  has {\it in effect} a boundary where it glues into the asymptotic AdS$_4$ geometry. The boundary is located in the asymptotic AdS$_2$  region shown by the dashed line at $r=r_c$ in Fig.(\ref{fig:geom}). 
 Fluctuations of this boundary are related to time reparametrisations and  determine the low-energy dynamics, at leading order. In order 
to be  glued into the AdS$_4$ region the near-horizon geometry must depart from the attractor AdS$_2$  solution with a constant value of the dilaton. The resulting variation of the dilaton  gives rise to an action for the fluctuations of the boundary which is determined by symmetry considerations to be the Schwarzian derivative of the time reparametrisations.  The coupling of the probe scalar  to the fluctuating boundary is also determined by symmetry considerations alone.  The leading order behaviour we found then arises from  the Schwarzian action coupled to the probe scalars in this way, just as in the JT model. 

The Einstein-Maxwell system does differ from the JT model in some important aspects. In the JT model the  geometry is  locally identical to  AdS$_2$ and only the dilaton departs from being a constant.
In contrast, in the Einstein-Maxwell system  the departure from the attractor solution   arises both due to the dilaton being non-constant and the geometry departing from AdS$_2$,
and both   these effects occur at linear order in the small parameter $L/r_h$. However, only the running of the dilaton is important, to leading order in $L/r_h$, in determining the low-energy response, since it is this effect, and not the departure of the geometry from AdS$_2$, which gives  rise to the Schwarzian action. Thus, differences between the JT model and the Einstein-Maxwell system become unimportant to leading order at low energies.

It will be worth checking how general our results are   and whether agreement with the JT model, and related  symmetry considerations, arises quite universally in near-extremal black holes.  In particular, it will be worth investigating whether this agreement arises in black holes in  other dimensions,
for black holes in asymptotically flat space, and importantly, beyond the S-wave sector. 

The microscopic models for extremal black holes in string theory are  different from the SYK model. In particular, they involve matrix degrees of freedom, or gauge groups with bi-fundamental matter which are similar to matrix degrees in their large $N$ behaviour. Our results show that at strong coupling these models must also exhibit the breaking of time reparametrisation symmetry and this breaking determines their low-energy dynamics. It will be worth checking if these results can be established directly by studying the large $N$  limit of these  models. 

The universality with which the JT model arises also motivates a further study of its properties. In particular, it would be worth  studying the quantum behaviour of this model in more detail. This can be done in the semi-classical limit by coupling the model to matter in a suitable large $N$ limit, which retains the effects of the quantum stress tensor of the matter while keeping gravity classical \cite{Almheiri:2014cka}. And  also more generally by attempting to quantize the full  theory including gravity. 

We hope to report on some of these directions in the future.

\acknowledgments
We thank Kristan Jensen, Gautam Mandal, Shiraz Minwalla, Subir Sachdev, Ashoke Sen and Spenta Wadia  for insightful discussions and comments.
We especially thank Juan Maldacena and Douglas Stanford  for their insightful comments, including important corrections in an earlier version of this paper.
PN acknowledges the support from the College of Arts and Sciences of the University of Kentucky.
AS would like to thank Juan Maldacena for an invitation to the Institute for Advanced Study, Princeton, where part of this work was done. 
RMS would like to thank the Perimeter Institute for hosting him under their ``Visiting Graduate Fellows'' program.
SPT thanks the organizers and the Simons Foundation for support to participate in the Simons Symposium on Quantum Entanglement (April 30th - May 6th, 2017) in  Krun, Germany, and acknowledges discussions  with the symposium participants.  He also thanks  the  organizers of the  ``It From Qubit" workshop (January 4th-6th, 2018) in Bariloche, Argentina, for their support and thanks  the workshop participants for discussion. 
SPT also acknowledges support from the J. C. Bose fellowship of the DST, Government of India. 
We thank the DAE, Government of India for support.  
We also acknowledge support from the Infosys Endowment for Research into the Quantum Structure of Spacetime. 
The work of AS is supported in part by the NSERC of Canada. 
Most of all, we thank the people of India for generously supporting research in String Theory.

 \section*{\Large Appendices}
\appendix

\section{Details of 4D Thermodynamic Calculations}
\label{fullthermo}
In this appendix, we give some details of the calculations involved in computing the partition function of AdS$_4$ RN black holes.
We start with the Euclidean Einstein-Maxwell action (including the counterterms),
\begin{align}
S_{reg}=-\frac{1}{16 \pi G}\int d^4 x\,\sqrt{g}\pqty{R-2\Lambda}&-\frac{1}{8\pi G}\int d^3 x\,\sqrt{\gamma}\,K\,+\,\frac{1}{4G}\int d^4 x \,\sqrt{g}\,F^2\nonumber\\
&+\frac{1}{4\pi G L}\int d^3 x\, \sqrt{\gamma}\,\pqty{1+\frac{L^2}{4}R_3},\label{ads4rnact2}
\end{align}
where $R_3$ is the Ricci scalar of the boundary surface. In the second line, we have added the appropriate counter terms to make the full 4D action finite, see \cite{Balasubramanian:1999re}.

We take the metric to be the AdS$_4$ RN black hole metric,
\begin{align}
ds^2&=a^2(r)\,dt^2+a^{-2}(r)\,dr^2,\label{ads4rnmet}\\
a^2(r)&=1-\frac{2GM}{r}+\frac{4\pi Q^2}{r^2}+\frac{r^2}{L^2},
\end{align}
and evaluate the action eq.\eqref{ads4rnact2}. The $r$ integral in the action goes from $r_+$, which is the location of the outer horizon, to $r_b$, which is the AdS$_4$ boundary. We get
\begin{align}
S_{bulk}&=-\frac{1}{16 \pi G}\int d^4 x\,\sqrt{g}\pqty{R-2\Lambda}\,+\,\frac{1}{4G}\int d^4 x \,\sqrt{g}\,F^2\nonumber\\
&=\frac{\beta}{2GL^2}(r_b^3-r_+^3)\,+\,\frac{2\pi Q^2 \beta}{G}\,\frac{1}{r_+}\label{ads4rnbulk},\\
S_{GH}&=-\frac{1}{8\pi G}\int d^3 x\,\sqrt{\gamma}\,K=\pqty{\frac{3M}{2}-\frac{r_b}{G}-\frac{3r_b^3}{2GL^2}}\beta\label{ads4rnbdy},\\
S_{ct}&=\frac{1}{4\pi G L}\int d^3 x\, \sqrt{\gamma}\,\pqty{1+\frac{L^2}{4}R_3}=\pqty{-M+\frac{r_b}{G}+\frac{r_b^3}{GL^2}}\beta\label{ads4rnct},\\
\Rightarrow S_{reg}&=S_{bulk}+S_{GH}+S_{ct}=\pqty{\frac M 2-\frac{r_+^3}{2GL^2}+\frac{2\pi Q^2}{G}\,\frac{1}{r_+}}\beta\label{ads4rnfull}.
\end{align}

To calculate the temperature, we take the near-horizon limit of \eqref{ads4rnmet},
\begin{align}
a^2(r_++\rho^2)&=a^2(r_+)\,+\,\rho^2\,\frac{da^2}{dr}\bigg|_{r_+}=\rho^2\,\frac{da^2}{dr}\bigg|_{r_+}\nonumber\\
&=\rho^2\pqty{\frac{2GM}{r_+^2}-\frac{8\pi Q^2}{r_+^3}+\frac{2r_+}{L^2}}\equiv \rho^2 a_2.\nonumber
\end{align}
The metric becomes
\begin{align}
ds^2&=\rho^2 \,a_2\,dt^2\,+\,\frac{4}{a_2}\,d\rho^2=\frac{4}{a_2}\pqty{\frac{a_2^2}{4}\rho^2\,dt^2\,+\,d\rho^2}\nonumber\\
\Rightarrow T&=\frac{a_2}{4\pi}=\frac{1}{2\pi}\pqty{\frac{GM}{r_+^2}-\frac{4\pi Q^2}{r_+^3}+\frac{r_+}{L^2}}\label{ads4rntemp}.
\end{align}

Now we can compute the entropy using the action \eqref{ads4rnfull} and \eqref{ads4rntemp}. We have
\begin{equation}
\label{ads4rnent}
S_{ent}=\beta M-S_{reg}=\pqty{\frac M 2+\frac{r_+^3}{2GL^2}-\frac{2\pi Q^2}{G}\,\frac{1}{r_+}}\beta=\frac{\pi r_+^2}{G}.
\end{equation}

\section{4D Calculation of the Four-Point Function}\label{genonact}
In this section we show some of the details involved in the computation of the on-shell action \eqref{isa}.
We start with a general background,
\begin{equation}
\label{swsol}
ds^2=a^2(r)\,dt^2\,+\,\frac{1}{a^2(r)}\,dr^2\,+\,b^2(r)(d\theta^2\,+\,\text{sin}^2\theta\,d\varphi^2),
\end{equation}
to which we add spherically symmetric perturbations described by \eqref{swpert}.
The stress energy tensor has components $T_{tt}$, $T_{rr}$, $T_{tr}$, $T_{\theta\theta}$ and $T_{\varphi\varphi}$, with each component being a function only of $t$ and $r$. Also, due to spherical symmetry we have $T_{\varphi\varphi}=T_{\theta\theta}\, \text{sin}^2\theta$.

The on-shell action with a probe scalar field is
\begin{align}
S&=\quarter \int d^4x \sqrt{g}\,\,\delta g^{\mu\nu}\,T_{\mu\nu}\nonumber\\
&=-\quarter \int d^4 x \sqrt{g}\,\pqty{\frac{1}{a^2}h_{tt}T_{tt}+\frac{1}{b^2}h_{\theta\theta}T_{\theta\theta}+\frac{1}{b^2\text{sin}^2\theta} h_{\theta\theta} T_{\varphi\varphi}}\nonumber\\
&=-\pi\int dt \, dr\,\pqty{\frac{b^2}{a^2}h_{tt}T_{tt}+2h_{\theta\theta}T_{\theta\theta}},\label{sonprob}
\end{align}
where in the last line we have integrated over $\theta,\varphi$. We have also made the gauge choice $h_{rr}=h_{tr}=0$ in writing the above expression. Let us manipulate the above action and write it in a form manifestly local in $r$ and only in terms of the stress tensor components by eliminating the metric perturbations.

The equations for the perturbations are
\begin{align}
\label{tteq}
&a^4\del_r^2 h_{\theta\theta}\,+\,a^4\left(\frac{a'}{a}+\frac{3b'}{b}\right)\del_r h_{\theta\theta}\,+\,\frac{a^2}{b^2}\left(1-\frac{8\pi Q^2}{b^2}\right)h_{\theta\theta}=8\pi G\, T_{tt},\\
\label{treq}
&\left(\frac{a'}{a}-\frac{b'}{b}\right)\del_t h_{\theta\theta}-\del_t\del_r h_{\theta\theta}=8\pi G \,T_{tr},\\
\label{rreq}
&\frac{1}{a^4}\del_t^2h_{\theta\theta}\,+\,\left(\frac{a'}{a}+\frac{b'}{b}\right)\del_r h_{\theta\theta}\,+\,\frac{b'}{b}\del_rh_{tt}\,+\,\frac{1}{a^2b^2}\left(1-\frac{8\pi Q^2}{b^2}\right)h_{\theta\theta}=8\pi G\,T_{rr},\\
&\frac{b^2}{a^2}\del_t^2h_{\theta\theta}\,+\,a^2b^2(\del_r^2h_{\theta\theta}+\del_r^2h_{tt})\,+\,2a^2b^2\left(\frac{a'}{a}+\frac{b'}{b}\right)\del_rh_{\theta\theta}\nonumber\\
\label{iieq}
&\hspace{30mm}+\,a^2b^2\left(\frac{3a'}{a}+\frac{b'}{b}\right)\del_rh_{tt}\,+\,\frac{16\pi Q^2}{b^2}h_{\theta\theta}=16\pi G\,T_{\theta\theta},
\end{align}
where a prime $'$ denotes a derivative w.r.t. $r$. 

The background equations relate $a$, $b$, $\Lambda$ and $Q$,
\begin{align}
\label{beq1}
&a^2b'^2-1\,+\,\Lambda b^2\,+\,2aba'b'\,+\,\frac{4\pi Q^2}{b^2}=0,\\
\label{beq2}
&a'^2b^2+\Lambda b^2\,+\,a^2b^2\left(\frac{a''}{a}+\frac{2a'b'}{ab}\right)-\frac{4\pi Q^2}{b^2}=0,\\
\label{beq3}
&b''=0.
\end{align}
Also, we have two conservation equations for the stress tensor,
\begin{align}
\label{cons1}
&\frac{1}{a^2}\del_tT_{tt}\,+\,a^2\del_rT_{tr}=-2a^2\left(\frac{a'}{a}+\frac{b'}{b}\right)T_{tr},\\
\label{cons2}
&\frac{1}{a^2}\del_tT_{tr}\,+\,a^2\del_rT_{rr}=-a^2\left(\frac{2b'}{b}+\frac{3a'}{a}\right)T_{rr}\,+\,\frac{a'}{a^3}T_{tt}\,+\,\frac{2b'}{b^3}T_{\theta\theta}.
\end{align}
Let us combine \eqref{beq1} and \eqref{beq2} to eliminate $\Lambda$ to find
\begin{equation}
b^{2} a^{'2} - a^{2} b^{'2}\,+\, a^{2} b^{2} \frac{a''}{a} + 1 - \frac{8\pi Q^{2}}{b^{2}} = 0.
\label{beq1.5}
\end{equation}
Now, we solve for $\partial_{r} h_{\theta\theta}$ using \eqref{treq},
\begin{equation}
\partial_{r} h_{\theta\theta} = \left( \frac{a'}{a} - \frac{b'}{b} \right) h_{\theta\theta} - \partial_{t}^{-1} \tau_{tr},
\label{drphi}
\end{equation}
where for convenience we have defined $\tau_{\mu\nu}$ as
\begin{equation}
\label{taudef}
\tau_{\mu\nu}=8\pi G T_{\mu\nu}.
\end{equation}
Plugging \eqref{drphi} into \eqref{rreq}, we find that $\partial_{r} h_{tt}$ is
\begin{equation}
\partial_{r} h_{tt} = \frac{b}{b'} \left[ \tau_{rr} - \frac{1}{a^4} \partial_{t}^{2} h_{\theta\theta} + \frac{a''}{a} h_{\theta\theta} + \left( \frac{a'}{a} + \frac{b'}{b} \right) \partial_{t}^{-1} \tau_{tr} \right].
\label{drhtt}
\end{equation}
Here, we've used the combined background equation \eqref{beq1.5} to simplify the coefficient of $h_{\theta\theta}$.
In terms of $\tau_{\mu\nu}$, the on-shell action \eqref{sonprob} is
\begin{equation}
S =- \frac{1}{8G}\int dt\, dr \left( \frac{b^{2}}{a^{2}} h_{tt} \tau_{tt} + 2 h_{\theta\theta} \tau_{\theta\theta} \right).
\label{on-shell-action-starting-pt}
\end{equation}
We can manipulate the first term as follows
\begin{align}
\int \frac{b^{2}}{a^{2}} h_{tt} \tau_{tt} &= \int \frac{b^{2}}{a^{2}} h_{tt} \partial_{t}^{-1} \partial_{t} \tau_{tt} \nonumber\\
&= - \int h_{tt} \partial_{t}^{-1} \partial_{r} \left\{ a^{2} b^{2} \tau_{tr} \right\} \quad \text{(using \eqref{cons1})} \nonumber\\
&= \int a^{2} b^{2} \partial_{r} h_{tt} \partial_{t}^{-1} \tau_{tr} \nonumber\\
&= \int a^{2} \frac{b^{3}}{b'} \left[ \tau_{rr} - \frac{1}{a^4} \partial_{t}^{2} h_{\theta\theta} + \frac{a''}{a} h_{\theta\theta} + \left( \frac{a'}{a} + \frac{b'}{b} \right) \partial_{t}^{-1} \tau_{tr} \right] \partial_{t}^{-1} \tau_{tr} \quad \text{(using \eqref{drhtt})}.\nonumber\\
\label{4dmanipulations1}
\end{align}
The first and last terms are manifestly local in $r$, so we can leave them as it is now.
We further manipulate the term involving $\partial_{t}^{2} h_{\theta\theta}$:
\begin{align}
- \int \frac{b^{3}}{a^{2} b'} \partial_{t}^{2} h_{\theta\theta} \partial_{t}^{-1} \tau_{tr} &= - \int \frac{b^{3}}{a^{2} b'} h_{\theta\theta} \partial_{t} \tau_{tr} \nonumber\\
&\!\!\!\!\!\!= \int h_{\theta\theta} \left\{ \partial_{r} \left( \frac{a^{2} b^{3}}{b'} \tau_{rr} \right) + \frac{a^{2} b^{3}}{b'} \left( \frac{a'}{a} - \frac{b'}{b} \right) \tau_{rr} - \frac{b^{3} a'}{a^{3} b'} \tau_{tt} - 2 \tau_{\theta\theta} \right\} \nonumber\\
&\quad\quad\quad\quad\quad\quad\quad\quad\quad \text{(using \eqref{cons2} and \eqref{beq3})} \nonumber\\
&\!\!\!\!\!\!= \int \frac{a^{2} b^{3}}{b'} \left\{ - \partial_{r} h_{\theta\theta} + \left( \frac{a'}{a} - \frac{b'}{b} \right) h_{\theta\theta} \right\} \tau_{rr} - 2 \int h_{\theta\theta} \tau_{\theta\theta} - \int \frac{b^{3} a'}{a^{3} b' } h_{\theta\theta} \tau_{tt} \nonumber\\
&\!\!\!\!\!\!=\int \frac{a^{2} b^{3}}{b'} \tau_{rr} \partial_{t}^{-1} \tau_{tr} - 2 \int h_{\theta\theta} \tau_{\theta\theta} - \int \frac{b^{3} a'}{a^{3} b' } h_{\theta\theta} \tau_{tt} \quad \text{(using \eqref{drphi})}.
\label{4dmanipulations2}
\end{align}
The first term is local in $r$ and the second term cancels the original $h_{\theta\theta} \tau_{\theta\theta}$ term in the action eq.\eqref{on-shell-action-starting-pt}, so we proceed only with the third term.
\begin{align}
- \int \frac{b^{3} a'}{a^{3} b'} h_{\theta\theta} \tau_{tt} &= - \int \frac{b^{3} a'}{a^{3} b'} h_{\theta\theta}  \partial_{t}^{-1} \partial_{t} \tau_{tt} \nonumber\\
&= \int \frac{b a'}{a b'} h_{\theta\theta} \partial_{t}^{-1} \partial_{r} \left( a^{2} b^{2} \tau_{tr} \right) \quad \text{(using \eqref{cons1})} \nonumber\\
&= - \int \frac{a b^{3} a'}{b'} \partial_{r} h_{\theta\theta} \partial_{t}^{-1} \tau_{tr} - \int  a^{2} b^{2} \partial_{r} \big(\frac{ba'}{ab'}\big) h_{\theta\theta} \partial_{t}^{-1} \tau_{tr} \nonumber\\
&=  \int a^{2} b^{2} \frac{ba'}{ab'} \left( \partial_{t}^{-1} \tau_{tr} \right)^{2} - \int \frac{a b^{3}}{b'} a'' h_{\theta\theta} \partial_{t}^{-1} \tau_{tr} \quad \text{(using \eqref{drphi} and \eqref{beq3})}.
\label{4dmanipulations3}
\end{align}
The second term here precisely cancels the corresponding term in the last line of \eqref{4dmanipulations1}, and so we are only left with manifestly $r$-local terms.
The final expression for the action becomes
\begin{equation}
\label{onshswact}
S=-8\pi^2 G\int dt\, dr\, \left(\frac{2a^2b^3}{b'}\,T_{rr}\frac{1}{\del_t}T_{tr}-a^2b^2\Big(1+\frac{2a'b}{b'a}\Big)\,T_{tr}\frac{1}{\del_t^2}T_{tr}\right),
\end{equation}
which agrees with eq.\eqref{isa}. 

In the region where the frequency dependence can be ignored, the solution for the probe field $\sigma$ is given by eq.\eqref{siga}. Such a form of the solution when substituted into eq.\eqref{onshswact} would give a contact term for the four-point function. To see this, consider the second term in the action eq.\eqref{onshswact} which is quadratic in $T_{tr}$,
\begin{equation}
\label{contactcalc}
\begin{split}
&\int dt\, T_{tr}\frac{1}{\del_t^2}T_{tr}\sim \int dt\,d\omega_1\,d\omega_2\,d\omega_3\,d\omega_4 \,\frac{\omega_1\,\omega_3}{(\omega_3+\omega_4)^2}\,\sigma(\omega_1)\sigma(\omega_2)\sigma(\omega_3)\sigma(\omega_4)\,\ e^{i\pqty{\omega_1+\omega_2+\omega_3+\omega_4}t}\\
&\sim \int dt_1\,dt_2\,\del_{t_1}\sigma(t_1)\,\sigma(t_2)\,\delta(t_1-t_2)\,\int dt_3\, dt_4\,d\omega_3\,d\omega_4\, \frac{\omega_3}{(\omega_3+\omega_4)^2}\,\sigma(t_3)\sigma(t_4)\,e^{i\omega_3(t_1-t_3)}\,e^{i\omega_4(t_1-t_4)},
\end{split}
\end{equation}
where we have used eq.\eqref{defsigo} and eq.\eqref{stress}. Note that we haven't been careful about keeping track of the $r$ dependence. Similarly, one can verify that we get an answer proportional to $\delta(t_1 -t_2)$ for the second term in the action eq.\eqref{onshswact} as well. Therefore, we see that the term in the action proportional to $\sigma(t_1)\sigma(t_2)\sigma(t_3)\sigma(t_4)$ will be non-zero only if $t_1=t_2$, which makes it a contact term.

\section{Relating the Sources at the AdS$_4$ Boundary and the Near-Horizon AdS$_2$ Boundary}
\label{4dto2d}
In this appendix, we will provide some details involved in relating the scalar sources at the AdS$_4$ screen and the asymptotic AdS$_2$ screen.

Since the scalar field satisfies a second order equation, two pieces of data are required to fix it. These are provided by the boundary condition eq.(\ref{ass2}) at the AdS$_4$ boundary, and the  horizon. For example, in Euclidean space, which we focus on here for concreteness,\footnote{The usual continuation to Minkowski time  gives the time-ordered Feynman correlators; however one can also obtain other correlators after a suitable analytic continuation from the Euclidean theory.}  the scalar field is regular at the horizon and does not blow up, and near the AdS$_4$ boundary behaves like eq.(\ref{ass2}).  Once  the solution is fixed, its value at $r=r_c$ can be  determined. 

In the asymptotic  AdS$_2$ region where $r_c$ is located, eq.\eqref{condrc1} and eq.(\ref{condrc2}), the scalar field can have  have two asymptotic behaviours,
\begin{equation}
\label{ass2a}
\sigma \rightarrow (r-r_h)^{{\tilde \Delta}_\pm}.
\end{equation}
Here ${\tilde \Delta_\pm}$ are the two characteristic fall-offs  towards the boundary, corresponding to the non-normalisable and normalisable modes for a field of mass $m$ in the AdS$_2$ spacetime,
\begin{equation}
\label{valtd}
{\tilde \Delta_{\pm}}={-1\pm \sqrt{1+ 4 m^2 L_2^2} \over 2}.
\end{equation}

A general solution will go like
\begin{equation}
\label{genassads2}
\sigma \rightarrow A \left[\pqty{{r-r_h\over \omega}} ^{{\tilde \Delta}_{+} }+ B \pqty{{r-r_h\over \omega}}^{\tilde \Delta_-}\right].
\end{equation}
$A$, the coefficient of the non-normalisable AdS$_2$ mode, acts effectively like the source in the near-horizon theory.  

Next we turn to the horizon. Regularity  at the horizon tells us that  the coefficient $B$ is independent of $\omega$. 
It then follows that the coefficient of the normalisable mode on the RHS of eq.(\ref{genassads2}) is suppressed compared to the coefficient of the non-normalisable mode by a factor of $\omega^{{\tilde \Delta}}$,
where
\begin{equation}
\label{deftd}
{\tilde \Delta}={{\tilde \Delta}_+}-{ {\tilde \Delta}_-}.
\end{equation}
At finite but small temperature, both temperature and frequency will enter in this ratio of coefficients in a combination whose overall power is still ${\tilde \Delta}$, so that the 
normalisable term above continues to be suppressed compared to the non-normalisable one. 

To make the suppression of the normalisable mode  manifest, we rewrite eq.(\ref{genassads2}) as 
\begin{equation}
\label{assads3}
\sigma \rightarrow C[(r-r_h) ^{{\tilde \Delta}_{+} }+ B\, \omega^{{\tilde \Delta}} (r-r_h)^{\tilde \Delta_-}],
\end{equation}
where 
\begin{eqnarray}
\label{defcd}
C&=& { A\over \omega^{{\tilde \Delta}_+}}.
\end{eqnarray}
Note that $C$ is $\omega$ dependent, while $B$ is $\omega$ independent. 

In evolving the solution eq.(\ref{assads3}) from the asymptotic AdS$_2$ region to the AdS$_4$ boundary the frequency term in eq.(\ref{preomf}) can be neglected. Therefore the contribution of the normalisable mode  can continue to be neglected for $r>r_c$. For purposes of determining the solution in this region we can therefore approximate eq.(\ref{assads3}) as 
\begin{equation}
\label{asscc}
\sigma \simeq C (r-r_h)^{{\tilde \Delta}_+}
\end{equation}
in the asymptotic AdS$_2$ region. 

Now suppose the non-normalisable mode in AdS$_2$, going like $(r-r_h)^{{\tilde \Delta}_+}$, in the asymptotic AdS$_4$ region becomes
\begin{equation}
\label{assads41}
(r-r_h)^{{\tilde {\Delta}}_+} \rightarrow  \alpha \left({ r \over L^2}\right)^{\Delta_+} +  \beta \left({r\over L^2}\right)^{\Delta_-},
\end{equation}
then using eq.\eqref{asscc} we get that 
\begin{equation}
\label{fass}
\sigma \rightarrow C \alpha \left({r\over L^2}\right)^{\Delta+}.
\end{equation}
Comparing with eq.(\ref{ass2}),  we see that $C$ is given in terms of the source term at the AdS$_4$ boundary  by 
\begin{equation}
\label{rels}
C(\omega)={\sigma(\omega)\over \alpha}.
\end{equation}

The multiplicative factor relating the two sources we talked about in section \ref{4ptfn}  is $\alpha$. This depends on the parameters $Q_m, L$ of the solution, which determine the interpolation for $r>r_c$,  but is independent of $\omega$. Once $C$ is fixed in terms of $\sigma(\omega)$ the solution in the region $r\le r_c$ is determined; in particular the asymptotic behaviour eq.(\ref{assads3}) is fixed, since the coefficient $B$  is determined by the scalar equation of motion and regularity at the horizon. 

The final result then is that for calculating the four-point function we can carry out the integral eq.(\ref{isa})  from the horizon to $r=r_c$ located in the asymptotic AdS$_2$ region, eq.(\ref{condrc1}) and eq.(\ref{condrc2}). 
The stress tensor in the integral is determined by the scalar field which satisfies the equation of motion and meets the boundary conditions of regularity at the horizon, and is of the form eq.(\ref{assads3})  in the asymptotic AdS$_2$ region, with $C$ given in terms of the source in the field theory by eq.(\ref{rels}).

\section{Thermodynamics of the JT Model}
\label{thermoap}
In this appendix, we give some details of the calculation of the on-shell action in the JT model with the metric eq.\eqref{JTBH}. 
The bulk term vanishes from the equation of motion and we are left with the following terms,
\begin{equation}
\label{JTonshell}
S_{JT}=-\frac{r_h^2}{4G}\pqty{\int d^2 x\,\sqrt{g} R+2\int_{bdy}\sqrt{\gamma}\,K}-\frac{r_{h}^{2}}{G}\int_{bdy}\sqrt{\gamma}\,\phi K+\frac{r_h^2}{GL_2}\int_{bdy}\sqrt{\gamma}\,\phi.
\end{equation}
Let us now evaluate each term of the action. The integral over $r$ runs from the horizon $r=r_h+\delta r_h$ to the boundary $r=r_c$.

The topological term in the action gives
\begin{align}
S_{top}&=-\frac{r_h^2}{4G}\pqty{\int d^2 x\,\sqrt{g} R+2\int_{bdy}\sqrt{\gamma}\,K}=-\frac{\beta\, \delta r_h\,r_h^2}{2GL_2^2} .\label{stop}
\end{align}
The contribution to the topological piece comes only from the horizon; the boundary terms cancel between the bulk and the extrinsic curvature term.

 To simplify the expression eq.\eqref{stop}, we relate $\delta r_h$ and $\beta$.
From eq.\eqref{ads4rntemp}, the temperature is
\begin{equation}
\label{temper}
T=\frac{1}{4\pi}\del_r(a^2)|_{r_+}=\frac{1}{4\pi}\,\frac{2\delta r_h}{L_2^2}.
\end{equation}

Therefore from eq.\eqref{temper} and eq.\eqref{stop}, we get
\begin{equation}
\label{stopf}
S_{top}=\frac{\pi r_h^2}{G}.
\end{equation}
We see that the topological piece gives the extremal entropy as expected.

Let us now compute the boundary term. We get
\begin{equation}
\label{sbdy}
S_{bdy}=-\frac{r_{h}^{2}}{G}\int_{bdy}\sqrt{\gamma}\,\phi K=-\pqty{\frac{\beta r_h}{GL_2^2}}(r_c-r_h)^2.
\end{equation}
This term is divergent and is canceled by adding the counter term eq.\eqref{Scounter}, which gives
\begin{align}
S_{count}&=\frac{r_h^2}{GL_2}\int_{bdy}\sqrt{\gamma}\,\phi\nonumber\\
&=\frac{\beta r_h}{GL_2^2}(r_c-r_h)^2-\beta \,\delta M\label{scount}.
\end{align}

Combining eq.\eqref{stopf}, eq.\eqref{sbdy} and eq.\eqref{scount}, we get the on-shell action
\begin{equation}
\label{JTactinter}
S_{JT}=-\beta \,\delta M-\frac{\pi r_h^2}{G}.
\end{equation}

\section{Coupling between the Probe Field $\sigma$ and the $\epsilon$ Modes for JT Model}
\label{secoup}
In this appendix, we derive the coupling between the time reparametrisation modes $\epsilon(t)$ and the probe scalar  field $\sigma$ in the JT model.  Consider the matter action for the JT model,
\begin{equation}
\label{jtact}
\begin{split}
S_{\sigma}&=2\pi r_h^2\int d^2 x \sqrt{g} \left((\del \sigma)^2+m^2\sigma^2\right)\\
&=-2\pi r_h^2\int_{\text{bdy}}\sigma\,\del_z\sigma.
\end{split}
\end{equation}
Near the boundary $\sigma(\omega, z)$ has a power law behaviour,
\begin{equation}
\label{bdysig}
\begin{split}
\sigma(\omega, z)&=\sigma(\omega)\,z^{\Delta_-}\,+\,C\, \omega^{\Delta_+-\Delta_-}\sigma(\omega)\,z^{\Delta_+}.
\end{split}
\end{equation}
Here $C$ is a constant. $\Delta_{\pm}$ are defined as follows,
\begin{equation}
\label{delpm}
\Delta_{\pm}=\frac{1}{2}\pm\sqrt{m^2L_2^2+\frac{1}{4}}.
\end{equation}
We impose Dirichlet boundary condition on $\sigma$ at the boundary $z=\delta$, and demand that at the boundary $\sigma= \sigma(\omega)\delta^{\Delta_-}$.
Therefore we get
\begin{align}
\sigma(\omega,z)&=\sigma(\omega)\,\delta^{\Delta_-}\pqty{\frac{z^{\Delta_-}+C\,\omega^{\Delta_+-\Delta_-}\,z^{\Delta_+}}{\delta^{\Delta_-}+C\,\omega^{\Delta_+-\Delta_-}\,\delta^{\Delta_+}}}\nonumber\\
&\simeq \sigma(\omega)(z^{\Delta_-}+C\,\omega^{\Delta_+-\Delta_-}\,z^{\Delta_+})(1-C\,\omega^{\Delta_+-\Delta_-}\,\delta^{\Delta_+-\Delta_-})\nonumber\\
\Rightarrow \sigma(t,z)&\simeq\sigma_-(t)\,z^{\Delta_-}+\,C\,(z^{\Delta_+}-z^{\Delta_-}\,\delta^{\Delta_+-\Delta_-})\,\sigma_+(t)\label{sigsol1},
\end{align}
where in going to the second line we Taylor expanded the denominator in powers of $\delta$. Here $\sigma_-(t)$ and $\sigma_+(t)$ are defined as
\begin{align}
\sigma_-(t)&=\int d\omega \,\sigma(\omega)\,e^{i\omega t}\label{sigm},\\
\sigma_+(t)&= c_1\int dt'\,\frac{\sigma_-(t')}{|t-t'|^{2\Delta_+}}\label{sigp}.
\end{align}
Here, 
\be
\label{defsigmp}
\sigma_+(t)={1\over 2 \pi} \int d\omega \,dt'\,\sigma_-(t') \,\omega^{\Delta_+-\Delta_-} \,e^{i\omega(t-t')},
\ee
and we take 
\be
\label{deft}
{1\over 2 \pi} \int d\omega \, e^{i \omega (t-t')} \, \omega^{\Delta_+-\Delta_-}= {c_1\over |t-t'|^{2 \Delta_+}},
\ee
where $c_1$ is a constant.
We now introduce fluctuations in the boundary by going to the new coordinate system $(\hat{t},\hat{z})$ defined via
\begin{align}
t&=\hat{t}+\epsilon(\hat{t}),\nonumber\\
z&=\hat{z}(1+\epsilon'(\hat{t}))\label{coordtrans}.
\end{align}
The boundary is now located at $\hat{z}=\delta$. The solution for $\sigma$, eq.\eqref{sigsol1}, becomes
\begin{align}
\hat{\sigma}(\hat{t},\hat{z})&=\hat{\sigma}_-(\hat{t})\,\hat{z}^{\Delta_-}\,+\,C\,\Big((1+\epsilon'(\hat{t}))^{\Delta_+}\,\hat{z}^{\Delta_+}-(1+\epsilon'(\hat{t}))^{\Delta_-}\,\hat{z}^{\Delta_-}\,\delta^{\Delta_+-\Delta_-}\Big)\,\hat{\sigma}_+(\hat{t})\nonumber\\
&\approx\hat{\sigma}_-(\hat{t})\,\hat{z}^{\Delta_-}\,+\,C\,\Big((1+\Delta_+\epsilon'(\hat{t}))\,\hat{z}^{\Delta_+}-(1+\Delta_-\epsilon'(\hat{t}))\,\hat{z}^{\Delta_-}\,\delta^{\Delta_+-\Delta_-}\Big)\,\hat{\sigma}_+(\hat{t})\label{sigsol2},
\end{align}
where in the second line we have expanded to linear order in $\epsilon(\hat{t})$. Here $\hat{\sigma}_-({\hat{t}})$ and $\hat{\sigma}_+({\hat{t}})$ are defined as follows,
\begin{align}
\hat{\sigma}_-({\hat{t}})&=\sigma_-(t(\hat{t}))\ (1+\epsilon'(\hat{t}))^{\Delta_-} ,\label{sighatm}\\
\hat{\sigma}_+({\hat{t}})&=c_1\int dt'\,\frac{\hat{\sigma}_-(t')(1+\epsilon'(t'))^{\Delta_+}}{|\hat{t}+\epsilon(\hat{t})-t'-\epsilon(t')|^{2\Delta_+}}\label{sighatp}.
\end{align}
The solution eq.\eqref{sigsol2} at the new boundary $\hat{z}=\delta$ becomes
\begin{equation}
\label{solbdy}
\hat{\sigma}(\hat{t},\delta)=\hat{\sigma}_-(\hat{t})\,\delta^{\Delta_-}\,+\,C\,(\Delta_+-\Delta_-)\,\epsilon'(\hat{t})\,\hat{\sigma}_+(\hat{t})\,\delta^{\Delta_+}.
\end{equation}
We have to impose the appropriate Dirichlet boundary condition for the new solution at the transformed boundary, namely, we require $\hat{\sigma}(\hat{t},\delta)= \hat{\sigma}_-(\hat{t})\,\delta^{\Delta_-}$. In order to do this, we note that $\hat{\sigma}$ is a solution to a linear differential equation and hence we can add a term $\delta\hat{\sigma}(\hat{t},\hat{z})=\delta\hat{\sigma}_-(\hat{t})\,\hat{z}^{\Delta_-}$ to eq.\eqref{sigsol2}, where $\delta\hat{\sigma}_-(\hat{t})$ is fixed by demanding that at the boundary $\hat{\sigma}$ has the correct boundary condition. Therefore we require,
\begin{align}
&\delta\hat{\sigma}_-(\hat{t})\,\delta^{\Delta_-}\,+\,C\,(\Delta_+-\Delta_-)\,\epsilon'(\hat{t})\,\hat{\sigma}_+(\hat{t})\,\delta^{\Delta_+}=0\nonumber\\
&\Rightarrow \delta\hat{\sigma}_-(\hat{t})=-C\,(\Delta_+-\Delta_-)\,\epsilon'(\hat{t})\,\hat{\sigma}_+(\hat{t})\,\delta^{\Delta_+-\Delta_-}\label{deltsigm}.
\end{align}
Note that  we should also add a term going like $\delta\hat{\sigma}_+(\hat{t})\,\hat{z}^{\Delta_+}$ to eq.(\ref{sigsol2})
to obtain a solution to the scalar equation meeting the regularity condition in the interior. However, this would give a sub-dominant contribution compared to $\delta\hat{\sigma}_-(\hat{t})\,\hat{z}^{\Delta_-}$ and we ignore it.

Adding $\delta\hat{\sigma}(\hat{t},\hat{z})$ to eq.\eqref{sigsol2}, and using eq.\eqref{deltsigm}, we get
\begin{align}
\hat{\sigma}(\hat{t},\hat{z})=\hat{\sigma}_-(\hat{t})\,\hat{z}^{\Delta_-}\,+\,C\,(1+\Delta_+\epsilon'(\hat{t}))\big(\hat{z}^{\Delta_+}-\hat{z}^{\Delta_-}\,\delta^{\Delta_+-\Delta_-}\big)\,\hat{\sigma}_+(\hat{t})\label{sigsol3}.
\end{align}

Now we compute the on-shell action by plugging eq.\eqref{sigsol3} into \eqref{jtact}. We get
\begin{align}
S_{\sigma}&=-2\pi r_h^2\int d\hat{t}\,\big(\hat{\sigma}_-(\hat{t})\, C\,(1+\Delta_+\epsilon'(\hat{t}))\,(\Delta_+-\Delta_-)\,\hat{\sigma}_+(\hat{t})\big)\nonumber\\
&=2\pi r_h^2\,C\,(\Delta_--\Delta_+)\int dt_1\,dt_2\,\frac{\hat{\sigma}_-(t_1)\,\hat{\sigma}_-(t_2)\,(1+\Delta_+\epsilon'(t_1))\,(1+\Delta_+\epsilon'(t_2))}{|t_1+\epsilon(t_1)-t_2-\epsilon(t_2)|^{2\Delta_+}}\label{matact1},
\end{align}
where in the first line we have dropped contact terms, and in the second line we have used eq.\eqref{sighatp}. Also we have absorbed $c_1$ into the constant $C$.

We now expand eq.\eqref{matact1} to linear order in $\epsilon(t)$ to get
\begin{equation}
\label{matact2}
S_{\sigma}=4\pi r_h^2\,C\,\Delta_+\,(\Delta_--\Delta_+)\int dt_1\,dt_2\,\frac{\hat{\sigma}(t_1)\,\hat{\sigma}(t_2)}{|t_1-t_2|^{2\Delta_+}}\pqty{\epsilon'(t_1)-2\,\frac{\epsilon(t_1)}{|t_1-t_2|}}.
\end{equation}

Let us now express the action eq.\eqref{matact2} in terms of the stress tensor components eq.\eqref{defT}. To the order we are working we can use the solution eq.\eqref{sigsol1}, where the boundary is not  fluctuating, to express the stress tensor components in terms of $\sigma$. The correction due the boundary fluctuations are higher order and will not  be relevant for the four-point function.

Therefore using eq.\eqref{sigsol1} and eq.\eqref{defT}, we get
\begin{align}
T_{zz}\Big|_{bdy}&=C\,(\Delta_+\Delta_--\Delta_-^2)\,\sigma_-(t)\,\frac 1 \delta \int dt' \frac{\sigma_-(t')}{|t-t'|^{2\Delta_+}},\label{Tzzbdy}\\
T_{tz}\Big|_{bdy}&=C\,(\Delta_+-\Delta_-)\,\sigma_-'(t)\int dt'\,\frac{\sigma_-(t')}{|t-t'|^{2\Delta_+}},\label{Tztbdy}
\end{align}
where we have again absorbed $c_1$ into $C$.
Now consider the quantity
\begin{equation}
\label{qty}
\int dt \,\pqty{\epsilon'(t)zT_{zz}+\epsilon(t)T_{tz}}.
\end{equation}
Expressing this in terms of $\sigma$, we get
\begin{align}
\int dt \,\pqty{\epsilon'(t)zT_{zz}+\epsilon(t)T_{tz}}&=\int dt\,\Bigg(C\,(\Delta_+\Delta_--\Delta_-^2)\,\epsilon'(t)\,\sigma_-(t)\int dt'\,\frac{ \sigma_-(t')}{|t-t'|^{2\Delta_+}}\nonumber\\
&+\,C\,(\Delta_+-\Delta_-)\,\epsilon(t)\,\sigma'_-(t)\int dt'\,\frac{ \sigma_-(t')}{|t-t'|^{2\Delta_+}}\Bigg).
\end{align}
We now do integration by parts in the second line to remove the derivative on $\sigma_-(t)$. After some simplifications, we get
\begin{equation}
\label{qty2}
\int dt \,\pqty{\epsilon'(t)zT_{zz}+\epsilon(t)T_{tz}}=C\Delta_+(\Delta_--\Delta_+)\int dt_1\,dt_2\,\frac{\sigma_-(t_1)\,\sigma_-(t_2)}{|t_1-t_2|^{2\Delta_+}}\pqty{\epsilon'(t_1)-2\,\frac{\epsilon(t_1)}{|t_1-t_2|}}.
\end{equation}
Comparing the expression eq.\eqref{qty2} with the action eq.\eqref{matact2} and ignoring corrections of $O(\epsilon^2)$, we get
\begin{equation}
\label{matact3}
S_{\sigma}=4\pi r_h^2\int dt \,\pqty{\epsilon'(t)zT_{zz}+\epsilon(t)T_{tz}}.
\end{equation}

\section{Dimensional Reduction with Sources}
\label{dimredsource}
In section \ref{dimred}, we studied the two-dimensional theory obtained by dimensional reduction of the action eq.\eqref{actem}. Let us now add a scalar field to the action and see how the solutions for the dilaton and metric perturbations get modified with a source.

We add the following term to the action eq.\eqref{actem},
\begin{equation}
\label{scalaract}
\frac{1}{2} \int d^4x \sqrt{\hat{g}} \, \Big((\nabla\sigma)^2 + m^2 \sigma^2\Big).
\end{equation}
We then perform the dimensional reduction of the complete action following the steps in section \ref{dimred}. The final bulk action quadratic in perturbations of the dilaton, eq.\eqref{flucdil}, and the perturbations of metric,
\begin{equation}
\label{metpert}
ds^2=\frac{L_2^2}{z^2}\left(\delta_{\alpha\beta}+h_{\alpha\beta}\right),
\end{equation}
 and with the source $\sigma$ is
\begin{equation}
\begin{split}
\label{qacte}
S_{bulk} = &- \frac{r_h^2}{4G} \int d^2x \sqrt{g}\, R - \frac{r_h^2}{2G} \int d^2x \, \phi \bigg[- \del^2 h+\del_\alpha \del_\beta h_{\alpha\beta}+\frac{\del_z h}{z}+ \frac{2 h_{zz}}{z^2}-\frac{2}{z} \, \del_\alpha h_{\alpha z}\bigg] \\
&+ \frac{3 r_h^2 \kappa}{G} \int d^2x \, \frac{\phi^2}{z^2}  -2\pi r_h^2\int d^2x \, h_{\alpha\beta} T^{\alpha\beta} + 4\pi r_h^2 \int d^2x \, \phi J,
\end{split}
\end{equation}
where the sources $T_{\alpha\beta}$ and $J$ are given by
\begin{align}
\label{stress2}
T_{\alpha\beta} &= \del_\alpha \sigma \del_\beta \sigma - \half \, \delta_{\alpha\beta} \bigg( \delta^{\mu\nu} \del_\mu\sigma \del_\nu\sigma + \frac{L_2^2}{z^2} m^2 \sigma^2\bigg),\\
\label{sj}
J &= \delta^{\mu\nu} \del_\mu\sigma \del_\nu\sigma + \half \frac{L_2^{2}}{z^2} m^2 \sigma^2.
\end{align}
Here, the stress tensor components satisfy the conservation equations
\begin{equation}
\label{stresscons}
\begin{split}
\del_t T_{tt}&=-\del_z T_{zt},\\
\del_t T_{zt}&=-\del_z T_{zz}-\frac{1}{z}(T_{tt}+T_{zz}).
\end{split}
\end{equation}
Varying $\phi$ in the action eq.\eqref{qacte}, we get an equation for $h_{\alpha\beta}$ which in the Fefferman-Graham gauge, $h_{tz} = h_{zz} = 0$, is
\begin{equation}
\label{htteq}
\del_z^2 h_{tt} - \frac{1}{z}\del_z h_{tt}=-12\kappa \frac{\phi}{z^2}-8\pi G J,
\end{equation}
where $\kappa$ is given in eq.\eqref{defkappa}; for a large black hole $\kappa =\frac 2 3$. 
Similarly, variation with respect to the metric fluctuations gives us the following equations for $\phi$,
\begin{equation}
\label{eomhtt}
\del_z^{2} \phi + \del_z\left(\frac{\phi}{z}\right) = 4\pi G\, T_{tt},
\end{equation}
\begin{equation}
\label{eomhzz}
\quad \del_t^2 \phi - \del_z\left(\frac{\phi}{z}\right) - \frac{2\phi}{z^2} = 4\pi G\, T_{zz},
\end{equation}
\begin{equation}
\label{eomhtz}
\quad \del_t\del_z\phi+ \del_t\left(\frac{\phi}{z}\right) = -\, 4\pi G\, T_{tz}.
\end{equation}
Eq.\eqref{eomhtt} can be solved to give
\begin{equation}
\phi = \frac{4 \pi G}{z} \int_{\infty}^{z} z' \int_{\infty}^{z'} T_{tt} (t,z'')\, dz'' dz' + C_{1} (t) z + \frac{C_{2} (t)}{z},
\label{phisoln-htt}
\end{equation}
where we have set the lower limit of integrations to $\infty$ by introducing two $z$ independent integration constants $C_1(t)$ and $C_2(t)$.  These  can be fixed using the constraint equations eq.\eqref{eomhzz} and eq.\eqref{eomhtz} to be of the form, 
\be
\label{constraintsf}
C_1(t)z+ { C_2(t)\over z}={d_0(t^2+z^2)+d_1t +c_2\over z}.
\ee
Requiring regularity at the horizon, $z=\infty$, and the initial condition that $\phi$ must take its background value at $t\rightarrow -\infty$, sets $d_0,d_1=0$, leading to 
\begin{equation}
\phi = \frac{4 \pi G}{z} \int_{\infty}^{z} z' \int_{\infty}^{z'} T_{tt} (t,z'')\, dz'' dz' + \frac{c_{2}}{z}.
\label{phi-soln-final}
\end{equation}
The value of $c_2$ can be fixed to be $\frac{L_2^2}{r_h}$ by examining the leading deviation of the dilaton from its unperturbed value, see eq.\eqref{valphi}.

Plugging in the solution for $\phi$, eq.\eqref{phi-soln-final}, into eq.\eqref{htteq} gives 
\begin{equation}
\label{htt-soln}
\begin{split}
h_{tt} = -48 \pi G \kappa &\int_{\infty}^{z} z_{1} \int_{\infty}^{z_{1}} \frac{1}{z_{2}^{4}} \int_{\infty}^{z_{2}} z_{3} \int_{\infty}^{z_{3}} T_{tt} (t,z_{4})\ dz_{4} dz_{3} dz_{2} dz_{1} \\ &-8\pi G\int_{\infty}^{z}z'\int_{\infty}^{z'}\frac{J(t, z'')}{z''}\,dz'\,dz''
- \frac{4 c_{2} \kappa}{z} + K_{1} (t) z^{2} + K_{2} (t).
\end{split}
\end{equation}
The integration constants $K_1(t)$, $K_2(t)$ set the integration limits to $\infty$. $K_{2}(t)$ can be done away by a scaling of time coordinate.
$K_{1}(t)$ corresponds to the time-reparametrisation modes discussed in section \ref{dimred}.

It will be useful to express the solution for the dilaton in a different way. We can write eq.\eqref{eomhtz} as
\begin{equation}
\label{eomhtz2}
\del_z\phi+ \frac{\phi}{z} = -\, 4\pi G\, \frac{1}{\del_t}\,T_{tz}.
\end{equation}
This equation should be thought of in $\omega$ space.

Combining eq.\eqref{eomhtz2} with eq.\eqref{eomhzz}, we get
\begin{equation}
\phi=\frac{4\pi G}{z}\pqty{z\frac{1}{\del_t^2}T_{zz}-\frac{1}{\del_t^3}T_{tz}}\,+\,\frac{L_2^2}{r_h z}.
\end{equation}
Here we have included the background solution for $\phi$, see eq.\eqref{phi-soln-final}. 
Once corrections to the metric are added, eq.(\ref{newmet}), eq.(\ref{valhtt}), to incorporate fluctuations in the boundary,  $\phi$ to the leading order in the stress tensor becomes
\begin{equation}
\label{phifluc}
\phi=\frac{4\pi G}{z}\pqty{z\frac{1}{\del_t^2}T_{zz}-\frac{1}{\del_t^3}T_{tz}}\,+\,\frac{L_2^2}{r_h z}(1-\epsilon'(t)).
\end{equation}

From eq.\eqref{phifluc}, $F[\sigma]$ defined in eq.\eqref{phi-soln-schematic} is given by
\begin{equation}
\label{fsig}
F[\sigma]=\frac{4\pi G}{z}\pqty{z\frac{1}{\del_t^2}T_{zz}-\frac{1}{\del_t^3}T_{tz}}.
\end{equation}

\bibliographystyle{JHEP}
\bibliography{refs}

\end{document}